\documentclass[10pt,conference]{IEEEtran} 

\usepackage{multirow}
\usepackage{graphicx}
\usepackage{amsmath}
\usepackage{color}
\usepackage{epsfig}
\usepackage{amsfonts}
\usepackage{amssymb}
\usepackage{amsthm}
\usepackage{epstopdf}
\usepackage{caption}
\usepackage{subcaption}
\usepackage{mathtools}
\usepackage{url}
\usepackage{bm}
\usepackage{hyperref}
\usepackage[ruled,vlined,linesnumbered,commentsnumbered]{algorithm2e}
\usepackage{mathtools}
\usepackage{epstopdf}
\usepackage{cite}
\usepackage{caption}
\usepackage{subcaption}
\usepackage{algpseudocode}
\usepackage{circledsteps}

\usepackage{tikz}
\newcommand*\circled[1]{\tikz[baseline=(char.base)]{
		\node[shape=circle,draw,inner sep=1pt] (char) {#1};}}

\usepackage{booktabs, tabularx}
\usepackage[referable]{threeparttablex}
\usepackage{stfloats}

\usepackage{setspace} 
\usepackage{float}

\newcommand\Tau{\mathcal{T}}
\newtheorem{Experiment}{Experiment}
\usepackage{setspace}

\usepackage{soul}   
\setstcolor{red}
\usepackage{xcolor} 

\usepackage{makecell}

\usepackage{caption}

\newcommand{\mbf}[1]{\mathbf{#1}}
\newcommand{\bols}[1]{\boldsymbol{#1}}
\newcommand{\tbf}[1]{\textbf{#1}}
\newcommand{\tit}[1]{\textit{#1}}

\begin{document}
\title{{\huge Learning a Common Dictionary for CSI Feedback in FDD Massive MU-MIMO-OFDM Systems}}
\author{Pavan Kumar Gadamsetty, \tit{Student Member, IEEE} and K. V. S. Hari, \tit{Fellow, IEEE} \\
 Lajos Hanzo, \tit{Life Fellow, IEEE} 
}
\maketitle

 \begin{abstract}

  In a transmit preprocessing aided frequency division duplex (FDD) massive multi-user (MU) multiple-input multiple-output (MIMO) {scheme assisted} orthogonal frequency-division multiplexing (OFDM) system, it is required to {feed back} the frequency domain channel transfer function (FDCHTF) of each subcarrier at the  user equipment (UE) to the base station (BS). The amount of channel state information (CSI)\footnote{We use the terms CSI and FDCHTF interchangeably} {to be fed} back to the BS increases linearly with the number of antennas and subcarriers, which may become excessive. Hence we propose a novel CSI feedback compression algorithm based on compressive sensing (CS) by designing a common dictionary (CD) to reduce the CSI feedback of existing algorithms. Most of the prior work on CSI feedback compression considered single-UE systems. Explicitly, we propose a common dictionary learning (CDL) framework for practical frequency-selective channels and design a CD {suitable} for {both} single-UE and multi-UE systems.  A set of two methods {is} proposed. Specifically, {the first one is} the CDL-K singular value decomposition (KSVD) method{, which} uses the K-SVD algorithm. The second {one is the} CDL-orthogonal Procrustes (OP) method, {which relies on solving} the orthogonal Procrustes problem. The CD conceived {for exploiting} the spatial correlation of channels across all the subcarriers and UEs  compresses the CSI at each UE, and {upon reception} reconstructs it at the BS. Our simulation results show that the proposed dictionary's estimated channel vectors have lower normalized mean-squared error (NMSE) than the traditional fixed Discrete Fourier Transform (DFT) {based} dictionary. The CSI feedback is reduced by 50\%, and the memory reduction at both the UE and BS starts from 50\% and increases with the number of subcarriers.

\end{abstract}


	\IEEEpeerreviewmaketitle

  \section{Introduction}

  Massive multiple-input multiple-output (MIMO) systems constitute a promising enabling technique for 5G/6G cellular networks as a benefit of their substantial spatial multiplexing gain~\cite{Sanguinetti_2018_IEEETWC_spatial_mux} in both time division duplex (TDD) and frequency division duplex (FDD) scenarios. At the base station (BS), combining the massive MIMO technology with orthogonal frequency-division multiplexing (OFDM) is capable of transmitting
  multiple data symbols to multiple UEs on the same time-frequency resource block, resulting in increased system throughput~\cite{Lajos_2015_IEEECOMSURV_50Years_MIMO}. In a multi-user (MU) massive MIMO-OFDM system, the knowledge of CSI is needed at the BS to implement transmit precoding (TPC) for suppressing the co-channel interference (CCI)~\cite{Liu_2019_IEEETWC_MU_Precoding}. In FDD systems, due to the absence of channel reciprocity~\cite{Ge_2020_GLOBECOM_Reciprocity}, the user equipment (UE) has to {feed back} the downlink (DL) frequency domain channel transfer function (FDCHTF) of each subcarrier to the BS. Feeding back the accurate CSI becomes more challenging with the increased number of antennas, subcarriers, and UEs~\cite{Lajos_2015_IEEEWCOM_Large_MIMO} .

  The compression of high-dimensional CSI is essential for reducing the CSI feedback. The wireless channels can be represented in a sparse form in the spatial-frequency domain using {`sparsifying'} bases termed as a dictionary~\cite{Gadamsetty_2023_NCC_Dictionary}. 
  In the compressive sensing (CS)-based feedback schemes~\cite{ozbek_2020_IEEEVTC_compressive}, the original CSI is mapped to a sparse domain using a dictionary. The traditional choice of the dictionary is a fixed Discrete Fourier Transform (DFT) matrix. Then a random Gaussian measurement matrix is introduced to compress the sparse vector {for feeding it back to the BS at a reduced rate}. The sparse signal is {then}  reconstructed at the BS using CS-based algorithms, such as {the} orthogonal matching pursuit (OMP)~\cite{Mo_2017_IEEETSP_OMP}, basis pursuit (BP)~\cite{Fu_2017_IEEETIT_Basis_Pursuit} or covariance-assisted matching pursuit (CAMP)~\cite{Xian_2019_IEEESPL_CAMP} {procedures}. The original CSI is then reconstructed by mapping the regenerated sparse signal back to the same dictionary used at the UE side.

  The authors of~\cite{Han_2015_GLOBECOM_basis} proposed a rotated version of the DFT basis to provide improved sparsity that results in reduced CSI mean-squared error (MSE) for a narrowband multi-user system with each UE having a single receive antenna. However, the proposed rotated basis does not exploit the antennas' spatial correlation. The massive MIMO-OFDM channel between a multi-antenna UE and the BS can be represented by a matrix~\cite{Lajos_2005_BOOK_ofdm_MU}. In such systems, the UE has to {feed back} the FDCHTF of each subcarrier, which results in huge feedback overhead. The FDCHTF feedback algorithm of a massive MIMO-OFDM single-UE system based on multidimensional compressive sensing theory using Tucker's tensor decomposition model is developed in~\cite{Chen_2014_VTC_Csi_ofdm}. Briefly, Tucker's tensor decomposition exploits the structure hidden in all the dimensions of the channel matrix and compresses it simultaneously in each dimension. The proposed scheme has a significant feedback reduction and hence improves the spectral efficiency. However, both the basis and measurement matrices should be learned. {The authors of~\cite{Wang_2021_IEEECOML_RLS_dictionary} introduced a recursive least squares dictionary learning algorithm (RLS-DLA) for CSI feedback. The proposed scheme achieves a substantial reduction in feedback requirements, however it requires the computation of large matrix inverses during the dictionary learning process. }

 
  \begin{table*}[t]
	\centering
	
	\caption{\label{tab:References_check} {COMPARING OUR CONTRIBUTION TO THE EXISTING LITERATURE}}
	\label{tab:widetable}
	\begin{tabular}{|l|l|l|l|l|l|l|l|}
		\hline 
		&{~\cite{Han_2015_GLOBECOM_basis}} & {~\cite{Wang_2021_IEEECOML_RLS_dictionary}} &{~\cite{ Li_2016_IEEECOML_Ref_KSVD}}&{~\cite{Jiang_2019_IEEETCOM_Low_rank}} & {~\cite{Prelcic_2020_IEEETWC_mmwave}}&{~\cite{Lau_2022_GLOBECOM_DL}} & Our\\ \hline
		Massive MIMO architecture &  \checkmark & \checkmark & \checkmark&\checkmark &\checkmark &\checkmark &\checkmark \\ \hline
		Spatially correlated channels & \checkmark  &\checkmark &\checkmark &\checkmark & & &\checkmark\\ \hline
		Sparsity &  \checkmark& \checkmark &\checkmark& &\checkmark &\checkmark &\checkmark\\ \hline
		Dictionary learning &   &\checkmark &\checkmark & &\checkmark & \checkmark&\checkmark\\ \hline
		Feedback savings & \checkmark & \checkmark &\checkmark &\checkmark &\checkmark &\checkmark &\checkmark\\ \hline
		Memory savings at UE  &  &  & & &\checkmark & &\checkmark \\ \hline
		Multi-user MIMO OFDM &  &  & & &\checkmark & &\checkmark\\ \hline
		UE mobility &  &  & & & & &\checkmark\\ \hline
		
	\end{tabular}
	\begin{tablenotes}[para] \footnotesize
		
	\end{tablenotes}
	
\end{table*}

  Another line of work focused on designing non-dictionary-based methods for FDCHTF feedback~\cite{Savazzi_2020_IEEEACC_CSI_Feedback}. In~\cite{Shim_2015_IEEETCOM_Antenna_group}, an antenna grouping-based method was proposed for reducing the feedback overhead by grouping multiple correlated antenna elements {into} a single representative value. By considering a ray-based channel model, the authors of~\cite{Heath_2018_IEEETCOM_AOD_codebook} and~\cite{Zhao_2021_IEEETVT_AOD_codebook} designed an angle-of-departure (AoD) based adaptive subspace codebook for feedback compression. In~\cite{Jiang_2019_IEEETCOM_Low_rank} and~\cite{Zhao_2021_IEEECOML_Low_rank} the authors exploited the low-rank characteristics of a large channel matrix for recovering the CSI at the BS.
  
  {Recent} solutions include Deep Learning (DL) techniques conceived for CSI compression and recovery using so-called {the} Bi-LSTM~\cite{LI_2019_IEEEACC_DL_CSI}, CsiNet-LSTM~\cite{Li_2019_IEEEWCL_CsiNet_LSTM}, DNNet~\cite{Li_2020_IEEECOML_DL_DNNet}, CS-ReNet~\cite{Li_2020_IEEETVT_DL_ReNet}, and DCRNet~\cite{Nallanathan_2022_IEEETVT_DCRNet} frameworks\footnote{For the expansion of these acronyms please refer to the relevant papers}. { Additionally, the application of Deep unfolding techniques has also shown promising results, as demonstrated in~\cite{Deyu_2023_IEEETWC_Deepunfold,Jin_2023_IEEEWCL_Deepunfold_bit}}. These techniques have better reconstruction performance than the {conventional} CS algorithms {of ~\cite{Ting_2012_WCNC_compressive}}, albeit at significantly increased computational complexity.

  {Massive MIMO-OFDM channels tend to be individually sparse and simultaneously share a common support set that} typically exhibit joint sparsity in the time domain (TD)~\cite{Vetterli_2012_IEEETCOM_Joint_sparsity}, which results in correlation among subcarriers in the frequency domain (FD). Since the DFT dictionary does not exploit spatial correlation across antenna arrays, we design a dictionary for massive MIMO-OFDM systems that can exploit the spatial correlation, hence achieving improved CSI reconstruction performance. The dictionary is generally learned from a training data set {by relying on} learning-based approaches~\cite{Li_2016_ICASSP_SBL,Cui_2018_IEEEWCL_Data_CL,Arslan_2019_IWCMC_DL_Beamspace,Prelcic_2020_IEEETWC_mmwave,Lau_2022_GLOBECOM_DL}. The dictionaries learned have the potential to offer improved normalized mean squared error (NMSE) performance compared to fixed dictionaries, like the DFT-based one. 
  {In}~\cite{Li_2016_IEEECOML_Ref_KSVD} {a} CS-based method was proposed, {which exploited} the spatial correlation among the antennas in a narrowband single-UE system using the K-SVD~\cite{Bruckstein_2006_IEEETSP_Ksvdalgo} algorithm. The method relies on learning the K-SVD dictionary from the training data set and on feeding back the K-SVD dictionary learned at the UE to the BS. Using this K-SVD dictionary, {the} CSI is compressed at the UE and reconstructed at the BS. The motivation for this K-SVD based dictionary is not only to reduce the CSI feedback, but also to reduce the NMSE {of CSI reconstruction}.

\begin{table}[]

		\caption{\label{tab:Acronyms} LIST OF ACRONYMS}
		\begin{tabular}{|l|r|}
			\hline

			\textbf{Acronyms} & \textbf{Meaning} \\ \hline
			\vspace{-3mm} 
		       & \\
			AOD & Angle-of-Departure \\ 
			BER & Bit Error Rate \\ 
			BS & Base Station \\ 
			CD & Common Dictionary\\ 
			CDL & Common Dictionary Learning \\
			CFR & Channel Frequency Response  \\
			CIR & Channel Impulse Response \\ 
			CS & Compressive Sensing \\
			CSI & Channel State Information\\ 
			DFT & Discrete Fourier Transform \\			
			DL  & Deep Learning \\ 
			FD & Frequency Domain \\
			FDD & Frequency Division Duplex \\
			FDCHTF  &  Frequency Domain Channel \\
			& Transfer Function  \\ 
			{GR-SVD} & Golub-Reinsch SVD \\
			K-SVD & K Singular Value Decomposition \\ 
			MIMO & Multiple-Input Multiple-Output \\
			MP & Matching Pursuit \\ 
			MSE &  Mean Squared Error \\  
			MU & Multi User \\ 
		    NMSE & Normalized Mean Squared Error \\ 
			OF & Objective Function \\ 	
			OFDM & Orthogonal Frequency Division \\& Multiplexing  \\
			OMP & Orthogonal Matching Pursuit \\ 
			OP  & Orthogonal Procrustes \\ 
		    QuaDRiGa & Quasi Deterministic Radio \\
			& Channel Generator  \\ 
			RA & Receive Antenna \\ 
			SU & Single User \\ 
			TA & Transmit Antenna \\
			TD & Time Domain\\ 
			TDD & Time Division Duplex\\ 
		    TPC & Transmit Precoding \\ 
			UE & User Equipment \\
			VQC & Vector Quantization Codebook \\  \hline

		\end{tabular}

\end{table}

  As the channel-induced dispersion is increased, the number of OFDM subcarriers also has to be increased to avoid an excessive performance degradation. Hence upon using the K-SVD algorithm of~\cite{Li_2016_IEEECOML_Ref_KSVD}, {the} number of subcarrier K-SVD based dictionaries increases as the number of subcarriers increases. Handling ubiquitous subcarrier K-SVD based dictionaries is cumbersome in terms of memory management and feedback load. To circumvent this problem, we propose a novel common dictionary learning (CDL) technique, which can replace the requirement of individual subcarrier K-SVD dictionaries, leading to the concept of a common dictionary (CD). {The CD effectively captures the channel characteristics of all the subcarriers and UEs, making it the optimal sparsifying dictionary for representing the channel's sparsity in massive MIMO systems. Given the CD learned, compressive channel estimation techniques can be constructed for acquiring the CSI}.
  {A set of two methods having different pros and cons are proposed for CDL, namely the CDL-KSVD method and the} CDL-orthogonal Procrustes (OP)~\cite{Dijksterhuis_2004_OxUni_procrustes} based method. These methods {are detailed} in Section III of the paper. Again, our primary motivation is to reduce {the} CSI feedback overhead on the uplink {as well as} the memory requirement at {both} the UE and the BS in FDD massive MU-MIMO-OFDM systems.

  
  \begin{table}[]
  	\caption{\label{tab:sysparameters_notations}LIST OF NOTATIONS}

  		\begin{tabular}{|l|r|}
  			\hline
  						\vspace{-3mm} 
  			& \\
  			$N_t (N_r)$ & Number of TAs (RAs)  \\  
  			$L$ & Number of channel taps in TD  \\  
  			$i$ & Index of the channel tap \\  
  			$N_c$  & Number of subcarriers in FD \\  
  			$l$  & Index of the subcarrier \\  
  			$K$  & Number of UEs   \\ 
  			$k $  &  Index of the UE \\ 
  			$V$  & Velocity of the UE \\
  			$S$  & Sparsity index \\  
  			$n$  & Frame index  \\  
  			$\bols{\Psi}$  & Dictionary \\  
  			$\bols{\Phi}$  & Measurement (Sensing) matrix \\  
  			$\bols{\Theta}$  & Product of a dictionary and 
  			$\bols{\Phi}$ \\ 
  			{$N_g$} & Number of rows in $\bols{\Phi}$ \\ 
  			{$g$} & Compression factor \\
  			${M}'$ & Total number of training channel vectors \\ 
  			$N$  & Number of training channel vectors \\ & considered across each subcarrier \\ 
 
            UE$_k$  & $k$-th user equipment \\  
            
            $\mbf{R}_{BS}$  & Spatial correlation matrix at the BS \\  
            $\mbf{R}_{UE,k}$  & Spatial correlation matrix at the $k$-th UE  \\ 
  			
  			$\mbf{\bar{H}}_{n,k}$  & TD channel matrix across the $n$-th tap \\ & of the $k$-th UE    \\  
            $\mbf{\breve{H}}_{n,k}^{}$ &  TD i.i.d channel matrix   \\  

  			$ {\mbf{H}}$  & Concatenated channel matrix of $K$ UEs   \\

    		$ \mbf{H}_k$  & Channel matrix of the $k$-th UE   \\
    	
    		$\hat{ \mbf{H}}_k$  & Reconstructed channel matrix of the $k$-th UE \\ & at the BS   \\
    
  			$ \mbf{H}'$  & Matrix of training channel vectors at the UE  \\ 
  			
  			$ \tilde{\mbf{H}}'$  & Sparse channel matrix representation of $ \mbf{H}'$ \\
  						
  			$ \hat{\mbf{{H}}}'$ &  Matrix of reconstructed training channel \\ & vectors at the BS\\ 
 
  			$ \hat{\tilde{\mbf{H}}}'$  &  Sparse channel matrix representation of $ \hat{\mbf{{H}}}'$ \\ 
  			
  			$\mbf{H}_{l_n,k}$  & Channel matrix or FDCHTF across the \\ & $l$-th subcarrier of the $k$-th UE in  $n$-th frame \\ 
  			
  			$\hat{\mbf{H}}_{l_n,k}$  & Reconstructed channel
  			matrix across the \\ &   $l$-th subcarrier of the $k$-th UE in $n$-th frame \\  
  			
  			${\mbf{h}}_{l_n,k}$ & Channel vector transformation of ${\mbf{H}}_{l_n,k}$ \\ 
  			
  			$\tilde{\mbf{h}}_{l_n,k}$  & Sparse channel vector representation of ${\mbf{h}}_{l_n,k}$\\ 
  			
  			$\hat{\mbf{h}}_{l_n,k}$  & Channel vector transformation of $\hat{\mbf{H}}_{l_n,k}$ \\ 
  			
  			$\hat {\tilde{\mbf{h}}}_{l_n,k}$  & Sparse channel vector representation of $\hat{\mbf{h}}_{l_n,k}$ \\ 
  			
  			$\mbf{h}_{c,l_n}^{k}$  & Compressed channel vector across the \\ & $l$-th subcarrier of the $k$-th UE in  $n$-th frame \\ 
  			  			
  			$\bols\Psi_{DFT}$ & DFT dictionary \\ 
  			
  			$\bols\Psi_{ksvd,l}^{k}$  & K-SVD dictionary across the $l$-th subcarrier \\ & of the $k$-th UE \\ 
  			
  			$\bols{\Psi}_{c}$ &  Common dictionary \\ 
  			  		  
  			$\Delta_{saved}$  & Total memory storage reduction  \\ 
  			
  			$\Upsilon$  & Dictionary feedback reduction factor \\
  			
  			$\Tau_{ksvd}$ & Feedback required for sending \\ & $N_c$ subcarrier dictionaries \\  
  			
  			$\Tau_{com}$ &  Feedback required for sending $\bols{\Psi}_c$ \\ 
  			
  			$\Tau_{saved}$  & Total dictionary feedback reduction  \\   \hline
  			 
  		\end{tabular}

  \end{table}

  \noindent {Main contributions of this article}:
  
  \begin{enumerate}
  \item We proposed a novel CDL framework for learning a CD, mainly using {the} CDL-KSVD and the CDL-OP methods. In the CDL framework {proposed} for a multi-UE system, the CD conceived {exploits} the spatial correlation of the {FD} channels across all the subcarriers and the UEs. {We demonstrate that this} implementation {improves} the NMSE performance when compared to the existing methods.

  \item In the CDL framework {proposed} for a single-UE system, the learning of CD is implemented at the UE. The UE sends only the CD to the BS in the uplink instead of all the subcarrier K-SVD dictionaries. This implementation reduces the dictionary feedback to the BS by a factor of $N_c$ and also reduces the memory requirement by having a single CD at the UE and the BS.	
  
  \item {We evaluate the proposed CD in the context of various system configurations and channel conditions in the face of UE mobility. The numerical results show a significant reduction in the NMSE of channel estimation and highlight the bit error rate (BER) performance of the channel estimates when using our learned dictionary. This corroborates the effectiveness of the CDL framework proposed over existing methods in wideband massive MIMO systems.}

  \end{enumerate}

  The remainder of this paper is organized as follows. Section II presents the system model, CS procedure, and the motivation. In Section III, the proposed methods are discussed. Then the application of {the} proposed methods in wideband systems is discussed in Section IV. Our simulation results are provided in Section V to show the NMSE performance of the proposed method compared to state-of-the-art methods. Finally, in Section VI, our conclusions are given.

  Notations: We use lower (upper) bold letters to denote column vectors (matrices) {and} super-scripts ${(.)^{-1}}, {(.)^{*}}, {(.)^H}$ to represent the inverse, complex conjugate and Hermitian operators respectively, $\|.\|_F $ denotes the Frobenius norm of a matrix; $\otimes$ denotes the Kronecker product, tr(.) is the
  trace of the matrix, vec(.) operation returns a column vector by stacking all the columns of a matrix.

	\begin{figure}[ht]
 \centering
\includegraphics[width=3.5in]{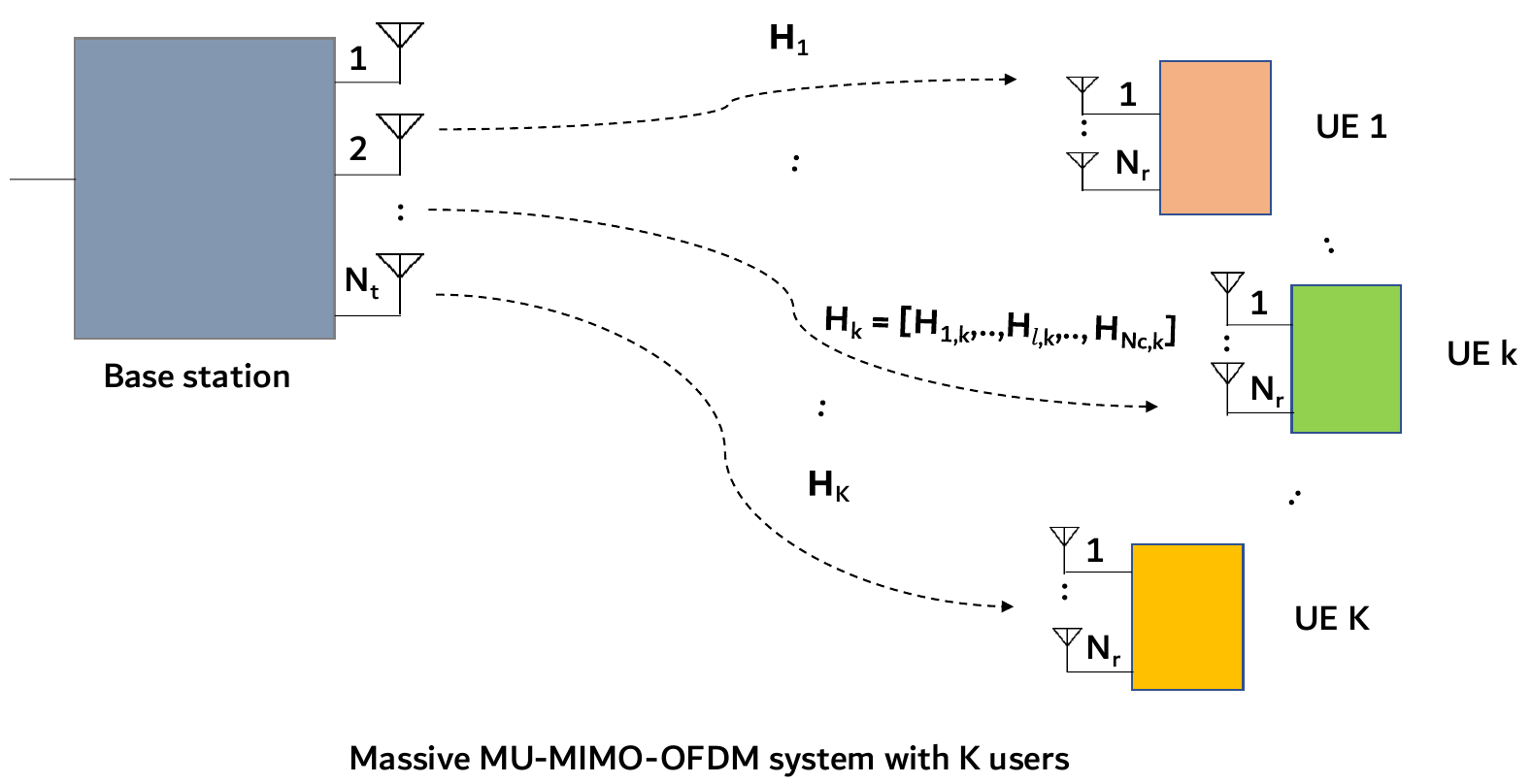}
		\caption{\footnotesize Overview of the considered massive MU-MIMO-OFDM system. Right: $K$ UEs  with $N_r$ RAs each for $k$ = $1$ to $K$; Left: massive MIMO base station with  $N_t$ TAs. $\mbf{H}_k$ represents the complete FDCHTF of the $k$-th UE and $\mbf{H}_{l,k}$ represents the  FDCHTF across the $l$-th subcarrier of the $k$-th UE. }
		\label{fig:massive_mimo_model}
	\end{figure}
	

\begin{table}[]
	\centering
	\caption{\label{tab:References_table} References for CSI compression techniques}
	\resizebox{8.8cm}{!}{
		\begin{tabular}{|l|l|l|}
			\hline
			& Single-UE & Multi-UE \\ \hline
			Narrowband & \begin{tabular}[c]{@{}l@{}} CS basis~\cite{Wang_2021_IEEECOML_RLS_dictionary},~\cite{Ting_2012_WCNC_compressive}\end{tabular} & \begin{tabular}[c]{@{}l@{}}Rotated DFT~\cite{Han_2015_GLOBECOM_basis}\\ K-SVD dictionary~\cite{Li_2016_IEEECOML_Ref_KSVD} \\ Bi-LSTM~\cite{LI_2019_IEEEACC_DL_CSI}\end{tabular} \\ \hline
			Wideband & \begin{tabular}[c]{@{}l@{}}Multidimensional CSI~\cite{Prelcic_2020_IEEETWC_mmwave}\\ CsiNet~\cite{Jin_2018_IEEEWCL_DL_CsiNet}\\ CsiNet-LSTM~\cite{Li_2019_IEEEWCL_CsiNet_LSTM}\\ DNNet~\cite{Li_2020_IEEECOML_DL_DNNet}\end{tabular} & CS-ReNet~\cite{Li_2020_IEEETVT_DL_ReNet} \\ \hline
	\end{tabular}}
\end{table}

	\section{System Model}

    In this section, we first introduce the massive MU-MIMO-OFDM channel and the associated spatial correlation matrices at the BS and UEs. Furthermore, we conceive the CS-based channel reconstruction procedure of massive MIMO channels. Next, we highlight the dictionary learning algorithms available in the literature. Then, in the final sub-section we describe the motivation of the proposed CDL framework. 
	
	\subsection{The Massive MU-MIMO-OFDM Channel}
	We consider a massive MU-MIMO-OFDM system using a uniform linear array (ULA) of $N_t$ TAs at the BS, $N_r$ RAs at all the $K$ UEs and $N_c$ subcarriers. For the $k$-th UE ($k$ = $1$ to $K$), consider a frequency-selective channel having $L$ taps in the TD. 
	Let $\mbf{H}_{l,k}$ represent the FDCHTF of the $l$-th subcarrier of the $k$-th UE given by
	\begin{equation}
		\mbf{H}_{l,k}=\sum\limits_{i=0}^{L-1}\mbf{\bar{H}}_{i,k}e^{-{j2\pi il\over N_c}},
	\end{equation}
	where $\mbf{\bar{H}}_{i,k} \in \mathbb{C}^{N_r \times N_t} $ is the $i$-th tap TD channel matrix. The tap coefficient $\mbf{\bar{H}}_{i,k}(p,q)$ represents the channel impulse response (CIR) of the link spanning from the $q$-th BS antenna to the $p$-th UE antenna.
	
	The spatial correlation of massive
	MIMO channels can be modeled by a Kronecker structure
	having separable transmit and receive correlation matrices~\cite{Ting_2012_WCNC_compressive},
	with $\mbf{\bar{H}}_{n,k}$ given by
	\begin{equation}
		\mbf{\bar{H}}_{i,k} = \frac{1}{\sqrt{\textrm{tr}(\mbf{R}_{UE,k}})}\mbf{R}_{UE,k}^\frac{1}{2}  \mbf{\breve{H}}_{i,k}^{}\mbf{R}_{BS}^\frac{1}{2}, 
		\label{Eq:correlated_chan}
	\end{equation}
	where $\mbf{\breve{H}}_{i,k}^{}$ is a $N_r \times N_t$ matrix whose elements are independent and identically distributed (i.i.d.) complex zero-mean, unit variance,
    Gaussian random entries. Furthermore, $\mbf{R}_{BS}$ and $\mbf{R}_{UE,k}$ are the spatial correlation matrices at the BS and $k$-th UE, respectively.
	
	The spatial correlation matrices are generated by Jakes' model often used in the literature, so the $uv$-th element of $\mbf{R}_{BS}$ and $\mbf{R}_{UE,k}$, can be modeled by
	$r_{uv}=  J_0 (\,{2\pi d_{uv}}/{\lambda})$
	where $d_{uv}$ is the distance between the antennas $u$ and $v$, $\lambda$ is the carrier wavelength and $J_0$(.) denotes the zeroth-order Bessel function of first kind~\cite{Albani_2015_IEEETAP_fading_chan_Bessel}.

\begin{figure*}[ht]
	\centering
	\scalebox{1}{%
		\begin{subfigure}[]{0.75\textwidth}
			\includegraphics[width=\textwidth]{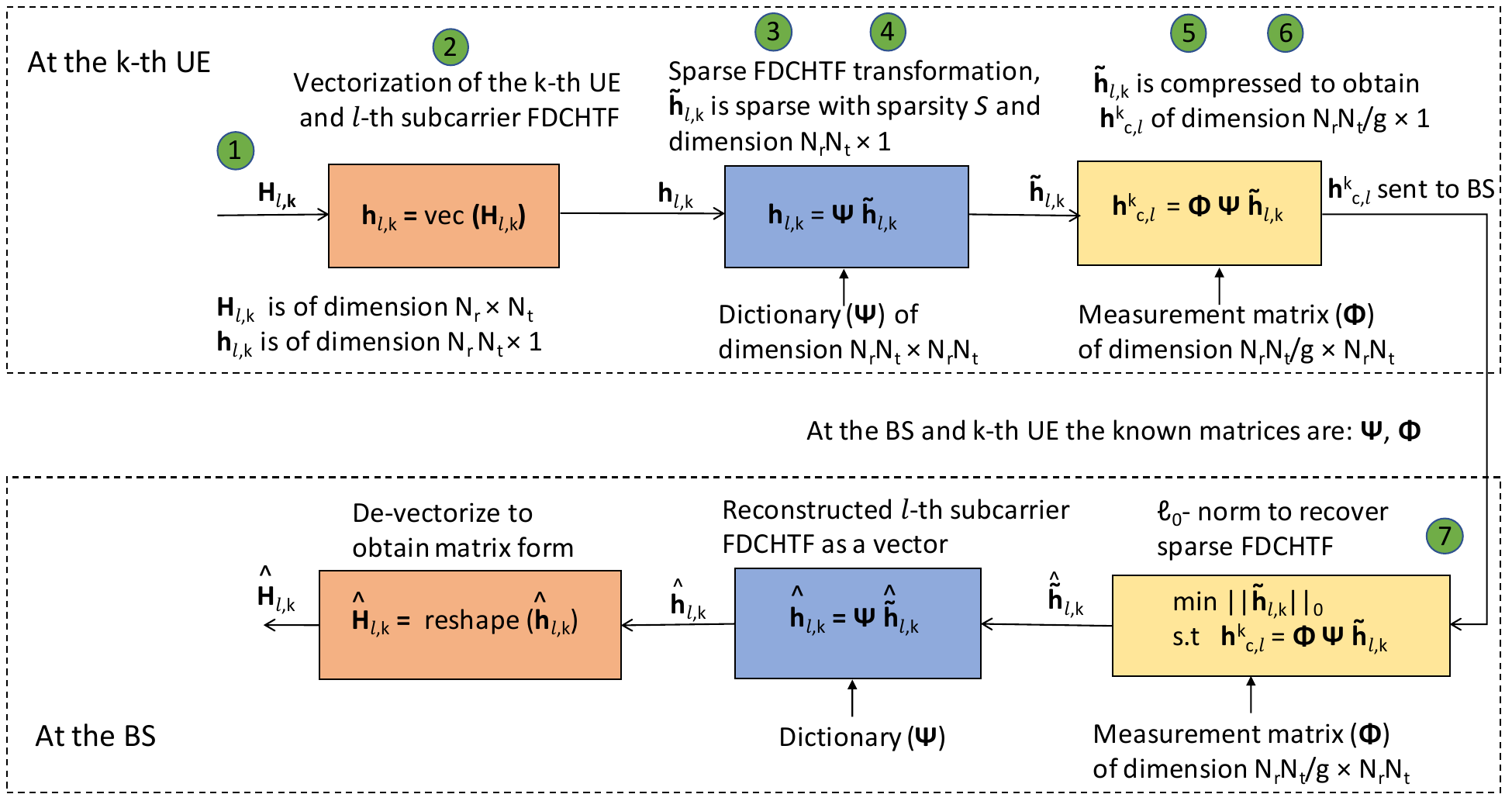}
		\end{subfigure}%
	}
	\caption{\footnotesize Overview of the considered CSI feedback compression scheme in the massive MU-MIMO-OFDM system. The FDCHTF across the $l$-th subcarrier of the $k$-th UE is compressed at the UE and reconstructed at the BS.}
	\label{fig:csi_compression}
\end{figure*}


\begin{figure*}[ht]
	\centering
	\scalebox{1}{%
		\begin{subfigure}[]{1\textwidth}
			\includegraphics[width=\linewidth]{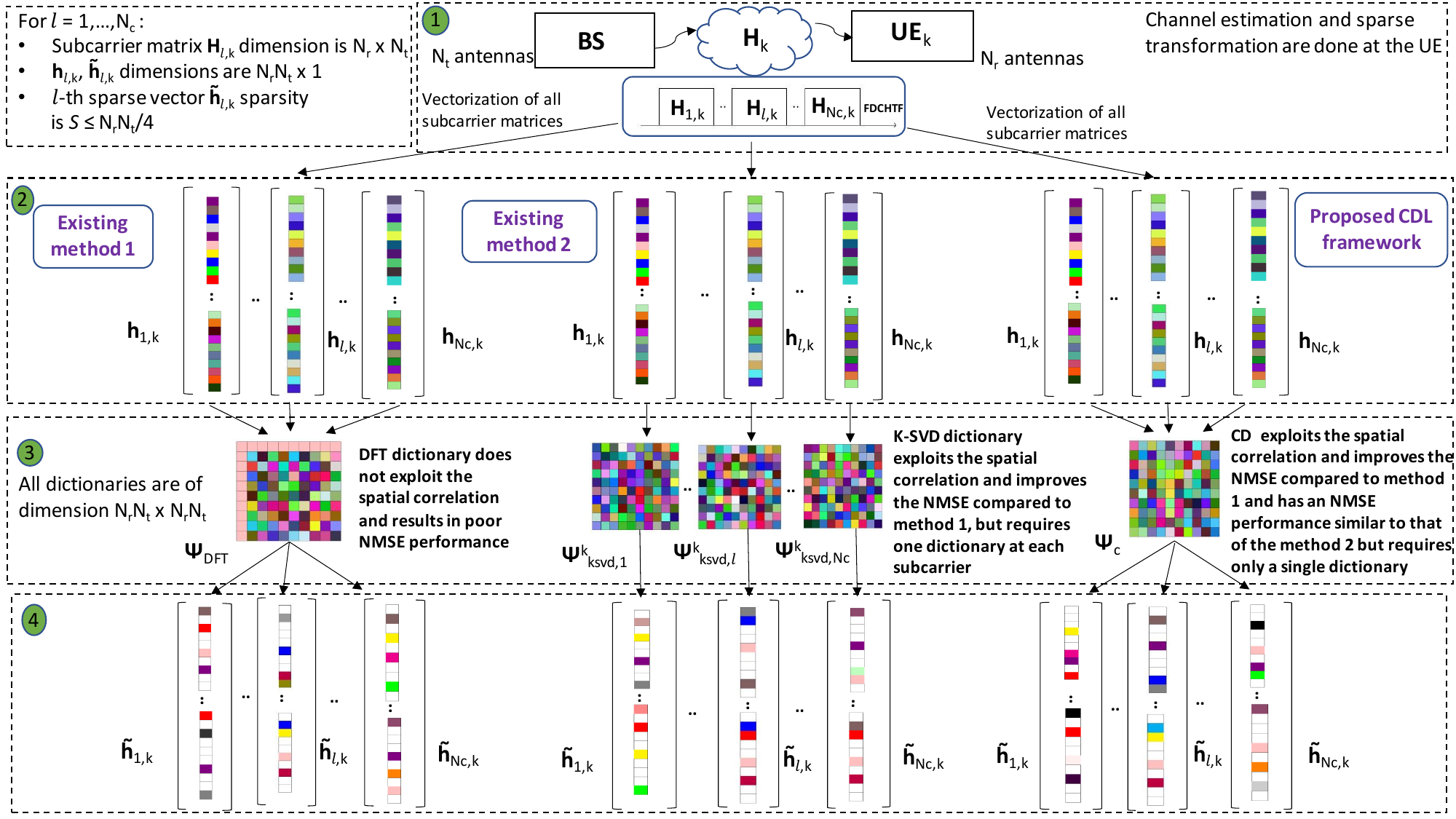}
		\end{subfigure}%
	}
	\caption{\footnotesize Overview of the massive MU-MIMO-OFDM system having a $k$-th UE with $N_r$ RAs and a BS with $N_t$ TAs. $\mbf{H}_k$ represents the complete FDCHTF at the $k$-th UE and $\mbf{H}_{l,k}$ represents the  FDCHTF across the $l$-th subcarrier of the $k$-th UE. {The} FDCHTFs are vectorized and then undergo sparse transformation using the dictionary obtained from the existing methods 1, 2 of~\cite{Li_2016_IEEECOML_Ref_KSVD} and the proposed framework. The circled numbers and the notations are the same as in Fig.\ref{fig:csi_compression}.  }
	\label{fig:massive_toy_model}
\end{figure*}


	\subsection{Compressive Sensing Based Channel Reconstruction}
	
	{Fig.~\ref{fig:csi_compression} shows the basic schematic of the FDCHTF compression and reconstruction across the $l$-th subcarrier of the $k$-th UE. More specifically, observe in Fig.~\ref{fig:massive_toy_model} that at the $k$-th UE we present the FDCHTF view  across all the subcarriers and its sparsification  using the existing as well as the proposed methods in parallel. The remainder of this section introduces each of the steps numbered in both figures.}
	
	\circled{1} In between the BS and a $k$-th UE the complete channel frequency response matrix that includes all the $N_c$ subcarrier channels is formed by stacking the channel matrices column-wise:     
	\begin{equation}
		\mbf{H}_k = [\mbf{H}_{1,k}, \;\hdots,\; \mbf{H}_{l,k},\; \hdots, \; \mbf{H}_{N_c,k}].
		\label{Eq:Channel_matrix}
	\end{equation}	

    \circled{2} We assume that the $k$-th UE perfectly estimates its channel matrix $\mbf{H}_k$, which should be shared with the BS through feedback. Instead of sending the FDCHTF of each subcarrier, the matrix $\mbf{H}_{l,k}$ is vectorized first into an $N_r N_t \times 1$ column vector using the vec(.) operation
	\begin{equation}
	      \mbf{h}_{l,k} = \textrm{vec}(\mbf{H}_{l,k}).
	      \label{Eq:vectorisation} 
	\end{equation}

	 \circled{3} \circled{4} In practical systems, the UE has to compress the estimated channel vector $\mbf{h}_{l,k} \in \mathbb{C}^{N_r N_t\times1}$ to avoid high feedback load.
    The wireless channel vector $\mbf{h}_{l,k}$ can be represented by a sparse vector~\cite{Ting_2012_WCNC_compressive} after a transformation 
    \begin{equation}
		\mbf{h}_{l,k} = {\bols{\Psi}} \tilde{\mbf{h}}_{l,k},
		\label{Eq:spar_1}
	\end{equation}
    where $\tilde{\mbf{h}}_{l,k}$ is the sparse representation of $\mbf{h}_{l,k}$. The number of non-zero components of a sparse channel vector is called the sparsity or sparsity index, and it is denoted by ${S}$,
    {while} ${\bols{\Psi}}$ is a $N_rN_t \times N_rN_t$ dictionary known to both the UE and the BS. A popular example of $\mbf{\Psi}$ is the DFT matrix. { Next, we introduce the measurement (sensing) matrix ${\bols{\Phi}}$, which plays a crucial role in compressive sensing. The measurement (sensing) matrix defines the measurement process in CS, which influences the reconstruction quality and efficiency of the signal recovery algorithm. It is responsible for mapping the original high-dimensional signal to a lower-dimensional signal. }
    
    \circled{5} \circled{6}  To compress the channel vector $\mbf{h}_{l,k}$, a measurement matrix ${\bols{\Phi}} \in \mathbb{C}^{N_g\times N_rN_t}$ $(N_g$\(<\!<\)$N_rN_t)$ that satisfies the Restricted Isometry Property (RIP)~\cite{Mo_2017_IEEETSP_OMP}, which {facilitates} sparse vector recovery {relying on:} 
	\begin{equation}
	\mbf{h}^{k}_{c,l} = {\bols{\Phi}}{\bols{\Psi}} \tilde{\mbf{h}}_{l,k},
	\label{Eq:spar_comp}
    \end{equation}
    where $\mbf{h}^{k}_{c,l}$ is the compressed channel vector {with dimension $N_g\times1$}. {Let us now} define $\bols{\Theta} = \bols{\Phi}\bols{\Psi}$.
    
    \circled{7} Then the reconstruction of $\mbf{h}_{l,k}$ can be
	formulated as an $\ell_0$-norm minimization problem and
	the sparse vector $\tilde{\mbf{h}}_{l,k}$ can be obtained by solving
	\begin{equation} 
		\min \limits _{\tilde{\mbf{h}}_{l,k}} \|{\tilde {\mbf{h}}_{l,k}}\|_{{0}} \quad s.t. ~~{{\mbf {h}^{k}_{c,l} }} = {\bols{\Theta}} {\tilde{\mbf{h}}_{l,k}}. 
		\label{Eq:spar_2}
	\end{equation}
    Thus, instead of feeding back $\mbf{h}_{l,k}$, the UE sends a low-dimensional vector $\mbf{h}^{k}_{c,l}$ to the BS for reducing the FDCHTF feedback. The BS reconstructs $\hat{\mbf{h}}_{l,k}$ from $\mbf{h}^{k}_{c,l}$, where $\hat{\mbf{h}}_{l,k}$ represents the reconstructed $\mbf{h}_{l,k}$. {The reconstructed channel vector $\hat{\mbf{h}}_{l,k}$ at the BS is utilized for precoding during the data transmission stage. The precoder matrices employed at the BS are denoted as $\mbf{W}^{g}_{}$ and $\mbf{W}^{g}_{op}$, which correspond to the beamforming weights. These weights are obtained from the true channels and the channels estimated using CDL-OP dictionary, respectively, for a compression factor of $g$. Here, the compression factor $g$ is defined as $g = \frac{N_rN_t}{N_g}$, where $N_g \times 1$ represents the dimension of the compressed channel vector $\mbf{h}^{k}_{c,l}$.}

	
	

\subsection{Motivation for the Common Dictionary Learning Framework}

  In CS-based feedback schemes, the traditional choice of the dictionary is a fixed DFT matrix, which does not exploit the spatial correlation between the antennas. The authors of~\cite{Han_2015_GLOBECOM_basis} proposed a rotated version of the DFT dictionary for better exploiting the sparsity, resulting in reduced FDCHTF mean-squared error (MSE) for a narrowband multi-user system supporting single antenna UEs. But this rotated basis still failed to exploit the antenna's spatial correlation for improving the MSE further.

  The authors of~\cite{Li_2016_IEEECOML_Ref_KSVD} have shown that a dictionary can be learned using the K-SVD algorithm for narrowband FDD massive SU-MIMO systems. This K-SVD dictionary {learned} exploits the spatial correlation between the antennas, and its FDCHTF reconstruction performance is improved compared to the fixed DFT dictionary. The proposed method requires FDCHTF and dictionary feedback to the BS. However, in practical communication systems, the channels are frequency-selective, and OFDM is a ubiquitous technique for such systems. In a massive MU-MIMO-OFDM system, to extend the idea of dictionary learning, it is {necessary} to feed back the FDCHTF and K-SVD based dictionary of each subcarrier of all UEs. Feeding back the entire FDCHTF $\mbf{H}_k$ of the $k$-th subcarrier will be a huge burden in the uplink. Another important issue is that substantial memory is required {for saving} all the $N_c$ subcarrier dictionaries at both the UE and the BS. The dimension of each subcarrier dictionary is $N_r N_t \times N_r N_t$, hence the memory required to store $N_c$ dictionaries is $N_c( N_r N_t)^2$.

  To overcome these challenges, we propose the novel idea of {a common dictionary}, which can replace the requirement of individual subcarrier dictionaries. The CD is designed for exploiting the spatial correlation across all the subcarriers and UEs in the FD, hence improving the CSI reconstruction accuracy.
  The proposed CD reduces the CSI feedback load and memory requirement in both single and multi-UE systems. In particular, the feedback load is further {reduced} for a single-UE system by sending only a single CD from the UE to the BS. 
  Hence, the proposed CDL framework reduces the CSI feedback and memory requirements, and we also study the NMSE performance compared to the DFT and subcarrier K-SVD dictionaries.


\begin{figure}[hbt!]
	\centering
	\includegraphics[width=2.75in]{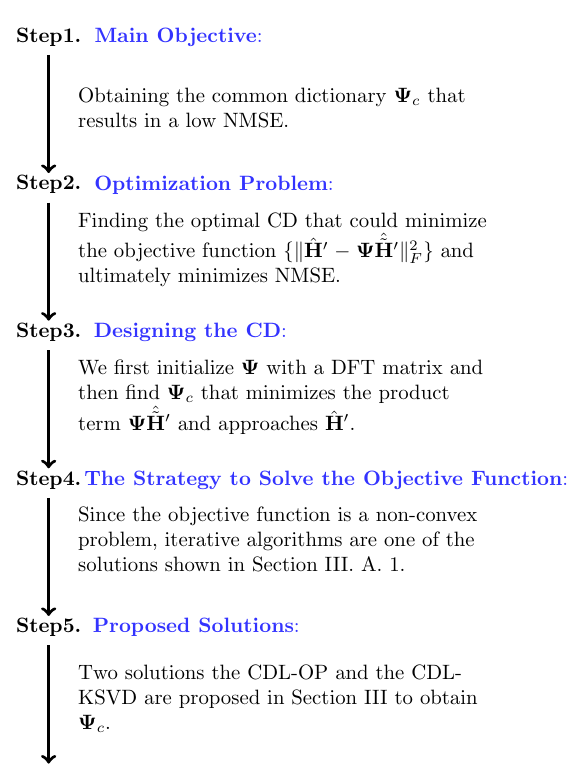}
	\caption{\footnotesize{ {Flow of the mathematical analysis.}}}
	\label{Fig:flowchart}
\end{figure}  



   \section{Proposed Common Dictionary Learning Framework}
   
   In this section, we detail the CDL framework proposed for a multi-UE system that constructs a dictionary from the estimated channel vectors and K-SVD based dictionaries of UEs.
   Before introducing the framework proposed, in Fig. ~\ref{Fig:flowchart} we provided a diagram showing the flow of the analysis described in the paper. This diagram guides the {reader} through the paper.

    \subsection{Common Dictionary Learning Framework}
    
    {The main goal of the proposed CDL framework is to construct a CD, denoted by $\bols{\Psi}_c$, that can exploit the correlation of channels across all the UEs and BS, for improving the CSI reconstruction at the BS. The matrix of training channel vectors is denoted by $\mbf{H}'$, which consists of $M'$ channel vectors collected for $N$ different frames across $N_c$ subcarriers and $K$ UEs. Then we have $M' = N \times N_c \times K$. }
    
    {To elaborate further, $\mbf{H}'$ is structured as $\mbf{{H}'} = [\mbf{H}'_1, \hdots, \mbf{H}'_k, \hdots , \mbf{H}'_K]$, and each sub-matrix in $\mbf{H}'$ is {represented} as $\mbf{H}'_k
    = [\mbf{H}'_{1,k}, \hdots, \mbf{H}'_{l,k}, \hdots, \mbf{H}'_{N_c,k}]$ $\forall k \in \{1, 2, \ldots, K\}$, $\forall l \in \{1, 2, \ldots, N_c\}$. Similarly, $\mbf{H}'_{l,k}$ is defined as $\mbf{H}'_{l,k}
    = [{\mbf{h}}_{l_1,k},\hdots,\mbf{h}_{l_n,k}, \hdots, {\mbf{h}}_{l_N,k}]$, where $\mbf{h}_{l_n,k}$ represents the channel vector transformation of $\mbf{H}_{l_n,k}$ at the $n$-th {MU-MIMO-OFDM} frame (time-instant). We assume that the channel envelope remains constant for {an OFDM} frame and then changes for the successive frames, 
    {according to the vehicular velocity}. Hence the consecutive frames are correlated.}  
    
    {The sparse representation of the matrix $\mbf{H}'$ is denoted as $\tilde{\mbf{H}}' = [\tilde{\mbf{H}}'_1, \hdots, \tilde{\mbf{H}}'_k, \hdots, \tilde{\mbf{H}}'_K]$,
    and each sub-matrix in $ \tilde{\mbf{H}}'$ is {represented} as 
    $\tilde{\mbf{H}}'_k
    = [\tilde{\mbf{H}}'_{1,k}, \hdots,\tilde{\mbf{H}}'_{l,k}, \hdots, \tilde{\mbf{H}}'_{N_c,k}]$ { $\forall k \in \{1, 2, \ldots, K\}$, $\forall l \in \{1, 2, \ldots, N_c\}$.} Similarly  $\tilde{\mbf{H}}'_{l,k} = [\tilde{\mbf{h}}_{l_1,k}, \hdots, \tilde{\mbf{h}}_{l_n,k},\hdots, \tilde{\mbf{h}}_{l_N,k}]$, where $\mbf{\tilde{h}}_{l_n,k}$ denotes the sparse representation of the channel vector $\mbf{h}_{l_n,k}$. }

    \noindent The CDL optimization problem is formulated as:
    \begin{align}
    & \min \limits _{\bols{\Psi}, \tilde {\mbf{H}}'} {\{ \|{\mbf{H}'} - {\bols{\Psi}} \tilde {\mbf{H}}' \|_{F}^{2}\}}  \quad \!\!\notag \\[0.5em]  \! \! & s.t.~ {\|\tilde {\mbf{h}}_{l_n,k}\|_{{0}}\leq } ~{S}, ~ \forall ~n \in \{1, 2, \ldots, N\}, 
     \quad \!\!\notag \\[0.5em] & ~ \forall ~l \in \{1, 2, \ldots, N_c\},  ~ \forall ~k \in \{1, 2, \ldots, K\}.    
    \label{Eq:CDL_Optimal}
    \end{align}    
    \noindent To solve the optimization problem in~\eqref{Eq:CDL_Optimal} we propose the following methods. 

    \subsubsection {CDL-KSVD method}
    
    \noindent {In the CDL-KSVD method the training set {$\mbf{H}'$} consists of the channel vectors of all the UEs.  The training set is employed to learn the dictionary {$\bols{\Psi}_c$} using the K-SVD algorithm\cite{Bruckstein_2006_IEEETSP_Ksvdalgo}. {The K-SVD algorithm has two stages: the sparse coding stage and the dictionary update stage. In the sparse coding stage, each column of ${\mbf{H}}'$ is sparsely represented using a dictionary. The dictionary update stage involves updating each column of $\bols{\Psi}$ with a dominant singular vector. As a result, the learned dictionary $\bols{\Psi}_c$ has unit-norm columns.} }

  	
    \begin{itemize}
  	
    	\item \noindent  {\em Sparse coding stage}: In the first stage of the K-SVD algorithm, the optimization problem is
    	formulated as:
    	\begin{align}
    		& \min \limits _{ \tilde {\mbf{H}}'} {\{ \|{\mbf{H}'} - {\bols{\Psi}} \tilde {\mbf{H}}' \|_{F}^{2}\}}  \quad \!\!\notag \\[0.5em]  \! \! & s.t.~ {\|\tilde {\mbf{h}}_{l_n,k}\|_{{0}}\leq } ~{S}, ~ \forall ~n \in \{1, 2, \ldots, N\}, 
    		\quad \!\!\notag \\[0.5em] & ~ \forall ~l \in \{1, 2, \ldots, N_c\},  ~ \forall ~k \in \{1, 2, \ldots, K\}. 
    		\quad   
    		\label{Eq:ksvd_obj_fun_sparse_upd}
    	\end{align}       
       {To solve~\eqref{Eq:ksvd_obj_fun_sparse_upd}, we begin by initializing the matrix $\bols{\Psi}$ with a DFT dictionary. The next step involves finding the matrix ${\mbf{H}}'$ having a sparse representation, which is an $\ell_0$ problem and it is carried out by using the OMP algorithm~\cite{Mo_2017_IEEETSP_OMP}. The objective of the OMP algorithm is to find a sparse representation of $ {\mbf{H}}'$ using a small number of non-zero elements in the matrix $\tilde {\mbf{H}}'$.}
    	
    	\item \noindent {\em Dictionary update stage}: In the second stage of the K-SVD algorithm,
        the optimization problem is formulated as:
    	\begin{align}
    		&\min \limits _{{\bols{\Psi}}}{\{ \|{\mbf{H}'} - {\bols{\Psi}} \tilde {\mbf{H}}' \|_{F}^{2}\}}. \quad  
    		\label{Eq:ksvd_obj_fun_dict_upd}
    	\end{align}
    	
    	{\noindent The solution to the problem posed in~\eqref{Eq:ksvd_obj_fun_dict_upd} is obtained by updating each column of the dictionary by computing a partial SVD of a matrix~\cite{Gadamsetty_2023_NCC_Dictionary}. After the dictionary update stage, the dictionary $\bols{\Psi}$ gets updated to $\bols{\mbf{\Psi}}_c$. It is to be noted that this method updates only the columns corresponding to sparse coefficients of the channel matrix $\tilde{\mbf{H}}'$.}  
    	
    	
     	\item { Repeat sparse coding and dictionary update stages until the stopping criterion is met}.
  	
    \end{itemize}

    The main advantage of the CDL-KSVD method is that it captures the spatial correlation of the channel vectors. But its drawback is that it requires a partial SVD operation for each column update in the dictionary update stage, which is computationally expensive.

   \subsubsection{CDL-OP method} 
   The CDL-OP method stands for common dictionary learning - orthogonal Procrustes method. In this method we solve the orthogonal Procrustes problem to learn the CD~\cite{Dijksterhuis_2004_OxUni_procrustes}. The CD obtained is a square matrix {with dimensions} $N_rN_t \times N_rN_t$. The constraint imposed is that the columns of the dictionary should be orthogonal. {The optimization problem} is formulated as: 
    \begin{align}
    	&\min \limits _{\bols{\Psi}} {\{ \|{\mbf{H}'} - {\bols{\Psi}} \tilde{\mbf{H}}' \|_{F}^{2}\}} \quad s.t.~~ \bols{\Psi}^H \bols{\Psi} =\mbf{I}, \quad 
    	\label{Eq:Orthogonal_procrust}
    \end{align}  
    where $\bols{\Psi}$ in~\eqref{Eq:Orthogonal_procrust} may be found explicitly by singular value decomposition (SVD).  
    \begin{align}
    \textrm{Let} \quad  &\mbf{C} =  \tilde{\mbf{H}}'  \mbf{H}'^H  \nonumber \\
    &[\mbf{U},\bols{\Sigma},\mbf{V}] = \textrm{SVD}(\mbf{C})  \nonumber \\
    &\bols{\Psi} = \mbf{V}  \mbf{U}^H.
    \label{Eq:Orthogonal_procrust_sol}
    \end{align}
    {The resulting dictionary $\bols{\Psi}$ obtained in~\eqref{Eq:Orthogonal_procrust_sol} is the CD ($\bols{\mbf{\Psi}}_c$). The main advantage of the CDL-OP method is that it captures the spatial correlation of the channel vectors, but at the cost of {an} SVD operation.}

%


\begin{figure*}[ht]
	\centering
	\includegraphics[width=4.5in]{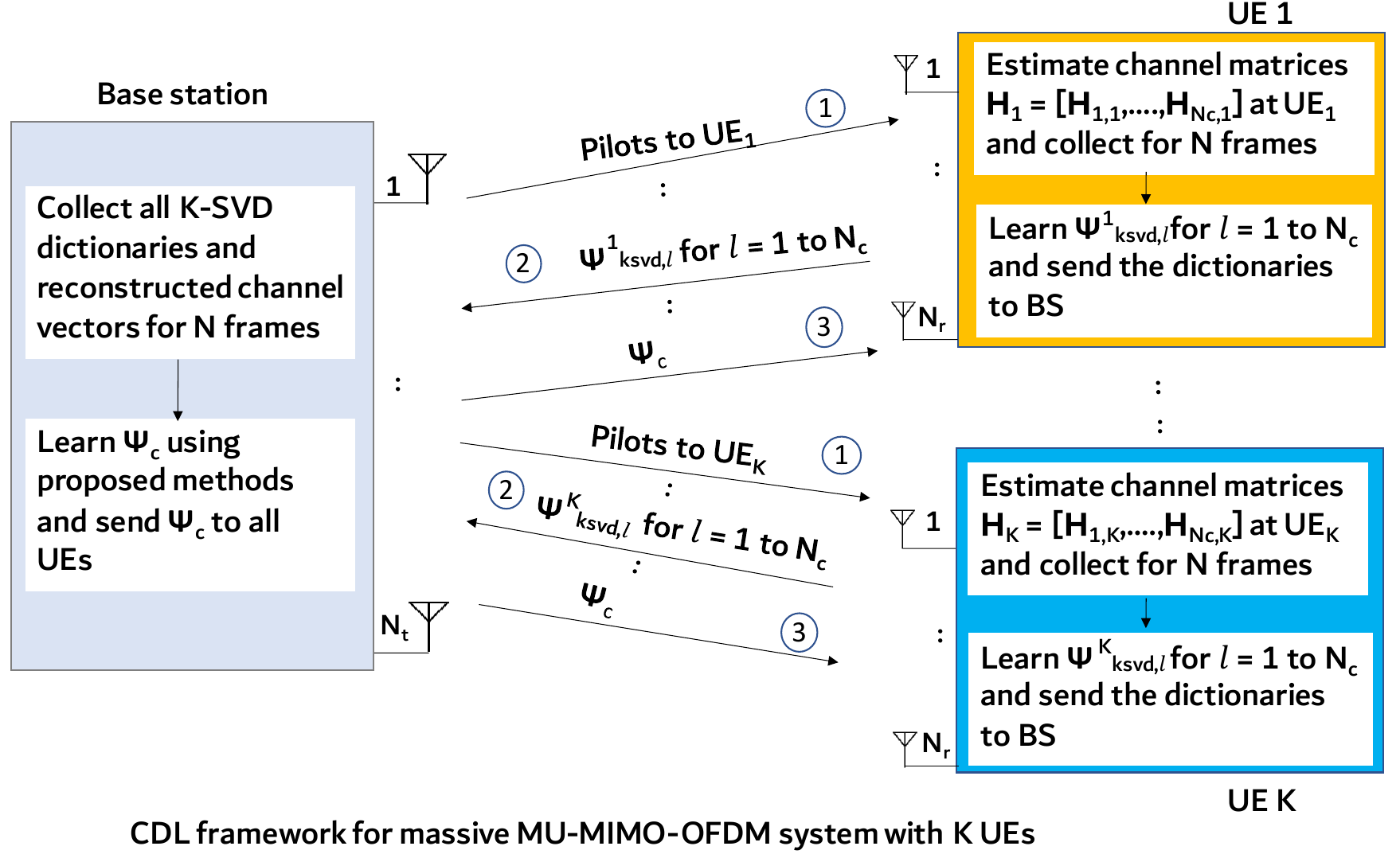}
	\caption{\footnotesize Overview of the CDL framework for FDD massive MU-MIMO-OFDM system. Right: $K$ UEs with $N_r$ receive-antennas each;  Left: massive MIMO base station with  $N_t$ transmit antennas; Center: For simplicity, only dictionary feedback is shown. }
	\label{fig:CDL_MU_MIMO}
\end{figure*}

 \subsection{Common Dictionary for Wideband Multi-UE System}

 A wideband channel has a broader signal bandwidth than the coherence bandwidth. In this section, we first discuss the CDL framework conceived for a wideband system supporting $K$ UEs, and then highlight the simplified scenario, where a single UE is present. Next, we will quantify the memory savings of storing only a single CD. Then in the final sub-section, we elaborate on the dictionary feedback reduction by sending only a {single} CD in a single-UE system.

  In the multi-UE system, since no communication takes place among the $K$ UEs, CDL is impossible at any UE. Consequently, the CDL is only feasible at the BS. For the CDL at the BS, we require the subcarrier dictionaries and the reconstructed channels (used as training channel vectors).  
  Let the reconstructed sparse channel matrix at the BS {be} represented by $\hat{\tilde{\mbf{H}}}' = [\hat{\tilde{\mbf{H}}}'_1, \hdots, \hat{\tilde{\mbf{H}}}'_k, \hdots, \hat{\tilde{\mbf{H}}}'_K]$, {where} each sub-matrix in $ \hat{\tilde{\mbf{H}}}'$ is represented as
  	$\hat{\tilde{\mbf{H}}}'_k
  	= [\hat{\tilde{\mbf{H}}}'_{1,k}, \hdots,\hat{\tilde{\mbf{H}}}'_{l,k}, \hdots, \hat{\tilde{\mbf{H}}}'_{N_c,k}]$ $\forall k \in \{1, 2, \ldots, K\}$, 
   {with} $\hat{\tilde{\mbf{H}}}'_{l,k} = [\hat{\tilde{\mbf{h}}}_{l_1,k}, \hdots, \hat{\tilde{\mbf{h}}}_{l_n,k}, \hdots,\hat{\tilde{\mbf{h}}}_{l_N,k}]$, {containing} $N$ reconstructed sparse channel vectors of each subcarrier, i.e., $\forall l \in \{1, 2, \ldots, N_c\}$. 
   
  Let the reconstructed matrix of training channel vectors at the BS be represented by $ \hat{\mbf{{H}}}' = [\hat{\mbf{H}}'_1, \hdots, \hat{\mbf{H}}'_k, \hdots , \hat{\mbf{H}}'_K]$, where each sub-matrix in $\hat{\mbf{{H}}}'$ is represented as $\hat{\mbf{H}}'_k
  = [\hat{\mbf{H}}'_{1,k}, \hdots, \hat{\mbf{H}}'_{l,k}, \hdots, \hat{\mbf{H}}'_{N_c,k}]$ $\forall k \in \{1, 2, \ldots, K\}$, with
  $\hat{\mbf{H}}'_{l,k} = [\hat{\mbf{h}}_{l_1,k}, \hdots,\hat{\mbf{h}}_{l_N,k}]$
  when considering $N$ reconstructed channel vectors for each subcarrier i.e., $\forall l \in \{1, 2, \ldots, N_c\}$.
  The total number of reconstructed training channel vectors in $\hat{\mbf{{H}}}'$ is $M' = N \times N_c \times K$.

 \noindent Importantly, at this stage we have to consider $\hat{\mbf{H}}'$ instead of $\mbf{H}'$ and $\hat{\tilde{\mbf{H}}}'$ instead of $\tilde{\mbf{H}}'$ in~\eqref{Eq:CDL_Optimal}.  The optimization problem of finding $\bols{\Psi}_c$ in the multi-UE system is formulated as follows:
 {
     \begin{align}
 	& \min \limits _{\bols{\Psi}, \hat{\tilde {\mbf{H}}}'} {\{ \|\hat{\mbf{H}}' - {\bols{\Psi}} \hat{\tilde {\mbf{H}}}' \|_{F}^{2}\}}  \quad \!\!\notag \\[0.5em]  \! \! & s.t.~ {\|\hat {\tilde{\mbf{h}}}_{l_n,k}\|_{{0}}\leq } ~{S}, ~ \forall ~n \in \{1, 2, \ldots, N\}, 
 	\quad \!\!\notag \\[0.5em] & ~ \forall ~l \in \{1, 2, \ldots, N_c\},  ~ \forall ~k \in \{1, 2, \ldots, K\}.    
 	\label{Eq:Optimal_multi}
 	\end{align}    }
\noindent The single-UE system ($K$ = 1) is a special case of a multi-UE system. For the current system when compared to the multi-UE system {of} Fig.~\ref{fig:CDL_MU_MIMO} there is no {need} to send the subcarrier dictionaries ($\bols{\Psi}^{k}_{ksvd,l}$, $\forall ~n \in \{1, 2, \ldots, N\},~ \forall ~l \in \{1, 2, \ldots, N_c\}$) from the UEs to the BS. The CD is learned at the UE itself using one of the two proposed methods and then the UE sends $\bols{\Psi}_c$ to the BS. Now {both the} BS and {the} UE will start using $\bols{\Psi}_c$. The total number of training channel vectors in $\mbf{H}'$ is $M' = N \times N_c$.


\noindent The optimization problem of finding $\bols{\Psi}_c$ is as follows:
{
\begin{align}
	&\min \limits _{\bols{\Psi},\tilde{\mbf{H}}'} {\{ \| \mbf{H}' - {\bols{\Psi}} \tilde{\mbf{H}}' \|_{F}^{2}\}} \quad \!\!\notag \\[0.5em]  \qquad & s.t.~ {\|\tilde {\mbf{h}}_{l_n}\|_{{0}}\leq } ~{S}, ~ \forall ~n \in \{1, 2, \ldots, N\},\!\!\notag \\[0.5em] & ~ \forall ~l \in \{1, 2, \ldots, N_c\}.\quad 
	\label{Eq:Optimal_single}
\end{align} }
\noindent The CDL framework of our multi-UE system is outlined in Algorithm~\ref{Algo:multiuser} and its block diagram is shown in Fig.~\ref{fig:CDL_MU_MIMO}. More specifically, in the Algorithm~\ref{Algo:multiuser} we present the step-by-step  procedure of the CDL framework. The remainder of this sub-section introduces each of the steps numbered in the Fig. \ref{fig:CDL_MU_MIMO} and the corresponding steps in the algorithm.

\circled{1} The BS sends the pilots to all the $K$-UEs in the system and each UE estimates its channels and follows the steps 4 to 10 in the algorithm for $N$ frames and learns the K-SVD based subcarrier dictionaries.

\circled{2} Each UE sends the $N_c$ K-SVD dictionaries to the BS. For the next $N$ frames each UE compresses the $l$-th subcarrier FDCHTF using the K-SVD dictionary $(\bols{\mbf{\Psi}}^k_{ksvd,l})$ and sends it to the BS. Then the BS reconstructs the $l$-th subcarrier FDCHTF with the aid of the same K-SVD dictionary. Using the reconstructed FDCHTFs and the K-SVD based subcarrier dictionaries of all the $K$ UEs, the BS learns the $\bols{\mbf{\Psi}}_c$. The \circled{2} procedure corresponds to the steps 14 to 19 in the algorithm.

\circled{3} After learning the CD at the BS, the BS sends the CD to all the $K$-UEs, the UEs and the BS will follow the steps 22 to 25 in the algorithm for FDCHTF compression and reconstruction using $\bols{\mbf{\Psi}}_c$.


\begin{algorithm}[h] 
	\DontPrintSemicolon
	\SetAlgoLined
	\SetKwInput{kwInit}{Init}
	\SetKw{Kw}{Variables}
	\KwIn{$\bols\Psi_{DFT}$, Pilots}  
	
	\KwOut{$\bols\Psi_{c}$}
	
	$\tbf{Variables}$: {Frame index $n$, UE index $k$, subcarrier index $l$. 
		
		$\textbf{Implementation}$: $\forall k \in \{1, \ldots, K\}$, $\forall l \in \{1, \ldots, N_c\}$}

	\For{$n$ = $1$ to $N$}     
	{			
		BS transmits pilot sequences for UE$_1$ to UE$_K$. \;
		
		The $k$-th UE estimates its channel matrix $\mbf{H}_{l_n,k}$ across $l$-th subcarrier from the pilots.\;
		
		After transformatation of $\mbf{H}_{l_n,k}$ to vector ${\mbf{h}_{l_n,k}}$, $k$-th UE at $l$-th subcarrier computes its compressed vector $\mbf{h}_{c,l_n}^{k}$ using ${\mbf{h}_{l_n,k}}$, $\tilde{\mbf{h}}_{l_n,k}$, $\bols\Psi_{DFT}$ and sends $\mbf{h}_{c,l_n}^{k}$ to BS. \;
		
		BS reconstructs $\hat{\tilde{\mbf{h}}}_{l_n,k}$ and $\hat{\mbf{h}}_{l_n,k}$ using $\bols\Psi_{DFT}$ and $\mbf{h}_{c,l_n}^{k}$. BS uses $\hat{\mbf{h}}_{l_n,k}$ for precoding. \;
		
		In parallel $k$-th UE at $l$-th subcarrier collects training channel vectors ${\mbf{h}_{l_n,k}}$'s to form matrix $[\mbf{h}_{l_1,k}, \hdots, \mbf{h}_{l_N,k}]$. 
		
		\If{$n$ = $N$}{	
			The $k$-th UE learns $\bols\Psi_{ksvd,l}^{k}$ from $[\mbf{h}_{l_1,k}, \hdots, \mbf{h}_{l_N,k}]$ and $\bols\Psi_{DFT}$ using K-SVD algorithm and sends $\bols\Psi_{ksvd,l}^{k}$(once) to BS.
		}	
	}

	\For{$n$ = $N+1$ to $2 N$}{
		Follow steps $4$ and $5$ for another $N$ frames. \;
		
		The $k$-th UE at $l$-th subcarrier computes compressed vector $\mbf{h}_{c,l_n}^{k}$ using $\mbf{h}_{l_n,k}$, $\tilde{\mbf{h}}_{l_n,k}$, $\bols\Psi_{ksvd,l}^{k}$ and sends $\mbf{h}_{c,l_n}^{k}$ to BS.\;
		
		BS reconstructs $\hat{\tilde{\mbf{h}}}_{l_n,k}$, $\hat{\mbf{h}}_{l_n,k}$ using $\bols\Psi_{ksvd,l}^{k}$, $\mbf{h}_{c,l_n}^{k}$.\;
		
		\If{$n$ = $2N$}{	
			BS computes $\bols\Psi_{c}$ using one of the two methods from $\hat{\mbf{h}}_{l_n,k}$'s, $\hat{\tilde{\mbf{h}}}_{l_n,k}$'s and $\bols\Psi_{ksvd,l}^{k}$ of all UEs and sends $\bols\Psi_{c}$ to all the UEs.\;
		}
		
	}
	
	\For{$i > 2 N$}{
		
		Follow steps $4$ and $5$ \; 
		
		The $k$-th UE starts using $\bols\Psi_{c}$ instead of $\bols\Psi_{ksvd,l}^{k}$ for step $15$ and sends compressed vector $\mbf{h}_{c,l_n}^{k}$ to BS.\;
		
		BS reconstructs $\hat{\tilde{\mbf{h}}}_{l_n,k}$ and $\hat{\mbf{h}}_{l_n,k}$ of $k$-th UE using $\bols\Psi_{c}$ and $\mbf{h}_{c,l_n}^{k}$(obtained from step $23$).\;
		
		BS uses $\hat{\mbf{h}}_{l_n,k}$ of each UE for precoding.\;
	}

	\caption{CDL Framework for $K$ Users}
	\label{Algo:multiuser}
\end{algorithm}


\subsubsection{\noindent Memory Reduction Calculation} 
\begin{itemize}
	\item The memory required to store each subcarrier dictionary $\bols{\Psi}^k_{ksvd,l}$ is $(N_r N_t)^2$, $\forall l \in \{1, 2, \ldots, N_c\}, \forall k \in \{1, 2, \ldots, K\}$.
	\item The total memory required to store $N_c$ subcarrier dictionaries is $N_c (N_r N_t)^2$. 
	\item The memory required to store $\bols{\Psi}_{c}$ is  $(N_r N_t)^2$.
	\item Total memory storage reduction for a $K$-UE system $=$ 
	$K \times$ ([Memory required to store $N_c$ subcarrier dictionaries] $-$ [Memory required to store $\bols{\Psi}_{c}]$), which is formulated as:
	\begin{align}
		\Delta_{saved} &= K(N_c -1) (N_r N_t)^2.
		\label{Eq:memory_MU}
	\end{align}
    \item The total memory storage reduction for a single-UE is
    \begin{align}    
    	\Delta_{saved} &=  (N_c -1) (N_r N_t)^2. 
    	\label{Eq:memory_SU}
    \end{align}

\end{itemize}


\subsubsection{\noindent Dictionary Feedback Reduction Calculation by Sending a CD in a Single-UE System}
\begin{itemize}
	\item The dimension of each subcarrier K-SVD dictionary $\bols{\Psi}_{ksvd,l}$ is $N_r N_t \times N_r N_t$, $\forall l \in \{1, 2, \ldots, N_c\}$.
	\item The total dimension of $N_c$ subcarrier dictionaries is $N_c \times N_r N_t \times N_r N_t$.
	\item The dimension of $\bols{\Psi}_{c}$ is $N_r N_t \times N_r N_t$.	
	\item Total dictionary feedback reduction for a single-UE ($\Tau_{saved}$) $=$ 
	[Feedback required for sending $N_c$ subcarrier dictionaries ($\Tau_{ksvd}$)] $-$ [Feedback required for sending $\bols{\Psi}_{c}$ ($\Tau_{com}$)], where we have    
	\begin{align*}
		& \Tau_{ksvd} = N_c (N_r N_t)^2 \quad \!\!\notag \\[0.5em]
		&\Tau_{com} =  (N_r N_t)^2 \quad \!\!\notag \\[0.5em] 
		&\Tau_{saved} = \Tau_{ksvd} - \Tau_{com}
		= (N_c-1) (N_r N_t)^2. \quad \!\!\notag 
	\end{align*}  
	\item {\em Reduction in dictionary feedback}: 
	We define the feedback reduction factor by ($\Upsilon$):
		\begin{equation}
			\Upsilon =  \displaystyle \frac{\Tau_{com}}{\Tau_{ksvd}} \\ =
			\displaystyle \frac{1}{N_c} 
		\label{Eq:feedback_SU}
	\end{equation}	
\end{itemize}

\begin{table}[H]
	\caption{Dictionary feedback reduction factor in a single-UE System.}
	\centering
	\begin{tabular}{l l l l l l}
		\hline
		$N_t$  & $N_r$  & $N_c$  & $\mathcal{T}_{ksvd}$ & $\mathcal{T}_{com}$ & $\Upsilon$\\ \hline
		$64$  & $1$  & $4$ & $16384$ & $4096$ & 
		$1/4$  \\ \hline
		$64$  & $1$  & $32$ & $131072$ & $4096$ & 
		$1/32$  \\ \hline
	\end{tabular}
	\label{Tab: Dictionary feedback_reduction_numbers}
	
\end{table}


\subsubsection{\noindent Computational Complexity of the Algorithm} 

{We calculate the computational complexity of the dictionary learning stage for both the CDL-OP and the CDL-KSVD methods.} 
 
\noindent {\em a) CDL-OP method}:
	\begin{itemize}
		\item {In~\eqref{Eq:Orthogonal_procrust_sol}, the SVD operation requires all the eigenvectors, resulting in a full SVD operation. 
		\item The computation of a full SVD operation, specifically using the Golub-Reinsch SVD (GR-SVD) method, requires 21$(N_rN_t)^3$ floating-point operations (FLOPS)~\cite{Gadamsetty_2023_NCC_Dictionary}. On the other hand, the Chan-SVD (R-SVD) method requires 26$(N_rN_t)^3$ FLOPS~\cite{Gadamsetty_2023_NCC_Dictionary}. 
		\item Therefore, to update the dictionary, we require a computational complexity of $\mathcal{O}[(N_rN_t)^3]$.}
	\end{itemize}

\noindent {\em b) CDL-KSVD method}:
	\begin{itemize} 
	 \item{In~\eqref{Eq:ksvd_obj_fun_dict_upd}, updating each column of the dictionary requires an SVD operation. This SVD operation only requires the dominant eigenvector, resulting in a partial SVD computation.
	 \item The GR-SVD method requires $14N_rN_tN_c'^2 + 9N_c'^3$ FLOPS~\cite{Gadamsetty_2023_NCC_Dictionary}, while the R-SVD method requires $6N_rN_tN_c'^2 + 20N_c'^3$ FLOPS~\cite{Gadamsetty_2023_NCC_Dictionary}. Here, $N_c'$ represents the number of non-zero coefficients corresponding to the $k$-th row in $\tilde {\mbf{H}}'$, and $N_c'$ ranges from 0 to $M'$. 
	 \item Therefore, to update the complete dictionary, a total of $N_rN_t$ partial SVD operations are required.}
	\end{itemize}
	 
\noindent c) {For example, let us consider $N_t = 64$, $N_r = 1$, $M' = 1600$, and an average value of $N_c' = M'/4$. The computational complexity in FLOPS is provided in Table~\ref{Tab:Computational_complexity_flops}. We represent the number of FLOPS required for updating a single column in the dictionary by CDL-KSVD (min), and that imposed by updating all columns in the dictionary using CDL-KSVD (max).}


\begin{table}[H]
	\caption{ {Computational complexity in FLOPS}}
	\centering
	\begin{tabular}{l l l }
		\hline
		Method  & GR-SVD  & Chan-SVD \\ \hline
		CDL-OP  & $5.505 * 10^6 $  & $6.8157 * 10^6$  \\ \hline
		CDL-KSVD (min)  & $7.1936 * 10^8$  & $1.3414*10^9$ \\ \hline
		CDL-KSVD (max)  & $4.6039*10^{10}$  & $8.5852*10^{10}$   \\ \hline
	\end{tabular}
	\label{Tab:Computational_complexity_flops}
	
\end{table}


\subsubsection{\noindent {CSI Feedback Case Study in a Single UE}}

{The dimension of the compressed channel vector $\mbf {h}^{k}_{c,l}$ ($\in \mathbb{C}^{N_g\times 1}$) sent from the UE in the uplink can be varied by adjusting the compression factor $g$. Specifically, the dimension is given by $N_g = \frac{N_rN_t}{g}$. By tuning the value of $g$, we can beneficially reduce the amount of CSI feedback required.}
	
{To quantify the feedback requirements, we introduce the variables $\gamma_{u}$ and $\gamma_{c}$ to represent the feedback for the non-dictionary and dictionary-based methods, respectively. In a non-dictionary based method without compression, the CSI fed back from the UE corresponds to  $N_cN_rN_t$ elements for one frame. For $N'$ frames the CSI feedback will be $\gamma_{u}$ = $N'N_cN_rN_t $. However, in a dictionary-based method associated with compression, the feedback is constituted by the CSI information having $\frac{N_cN_rN_t}{g}$ elements for one frame, along with a one-time transmission of a dictionary with $(N_rN_t)^2$ elements. So, for a total of $N'$ frames the feedback is given by $\gamma_{c}$ = $(N_rN_t)^2$ + $\frac{N'N_cN_rN_t}{g}$.}

{We define $\Gamma$ as the CSI feedback ratio, which is calculated as the ratio between ${\gamma_c}$ and ${\gamma_u}$. If $\Gamma<1$, it indicates that the value of $\gamma_c$ is lower than $\gamma_u$, resulting in a saving in CSI feedback.}
\begin{equation}
	{\Gamma  =  \frac{\gamma_c}{\gamma_u} 
	       =  \frac{(N_rN_t)^2 + \frac{N'N_cN_rN_t}{g}} {N'N_cN_rN_t}	
	       = \frac{N_rN_t}{N'N_c}+\frac{1}{g} 
	       \label{Eq:CSI_feedback_Savings_SU}	}       
\end{equation}

{For example in Table~\ref{Tab: Feedback_comparison_table_in_numbers} we consider a scenario associated with $N'=2^{10}$ and vary the values of $N_c$, $N_r$, and $g$. Using the formula given in \ref{Eq:CSI_feedback_Savings_SU}, we demonstrate significant reductions in CSI feedback. The level of compression applied to the channel vector $\mbf {h}^{k}_{c,l}$ depends on the sparsity parameter $S$. As per the lower bound, the number of elements in $\mbf {h}^{k}_{c,l}$ must satisfy $N_g > 2S$~\cite{Fu_2017_IEEETIT_Basis_Pursuit}. In Fig. \ref{Fig:BER_coded_quad_64A_16Psk}, we illustrate the impact of the compression factor $g$ on the BER vs. signal-to-noise ratio (SNR) performance for both the non-dictionary and dictionary-based methods.}

\begin{table}[H]
	\caption{{CSI feedback savings comparison table for $N'$ = $2^{10}$.}}
	\centering
	\begin{tabular}{l l l l l  }
			\hline
			$g$ & $N_c$  & $N_t$ & $N_r$ & $\Gamma$ = $\frac{\gamma_{c}}  {\gamma_{u}}$  \\  \hline 	
			
			2 &  32  & 64 & 1   &  $0.5$  \\ \hline
			2 &  64  & 64 & 2   & $0.5$  \\ \hline
			4 &  32  & 64 & 1   & 0.252 \\ \hline
			4 &  64  & 64 & 2   & 0.252   \\ \hline
			
		\end{tabular}
	\label{Tab: Feedback_comparison_table_in_numbers}
	
\end{table}

\section{Numerical Results}   
 
In this section, we {provide} the simulation results for characterising the NMSE performances of the DFT dictionary ($\bols{\Psi}_{DFT}$), {of the} individual K-SVD based subcarrier dictionaries ($\bols{\Psi}^{k}_{ksvd,l}$) and {of the} proposed CD ($\bols{\Psi}_{c}$). Using the DFT dictionary as the initial reference dictionary, we obtain the individual K-SVD based subcarrier dictionaries. Then $\bols{\Psi}_{c}$ is learned by {the} proposed methods. 
For NMSE calculations, each dictionary is used for reconstructing the $P$ channel vectors of each subcarrier at the BS.

\noindent The NMSE of the reconstructed channel is used as a performance metric defined as 
\begin{equation}
	\mbf{NMSE} = \frac{1}{P} \smashoperator[r]{\sum_{i=1}^{P}} \frac{\|\hat{\mbf{h}}_{l_n,k}-\mbf{h}_{l_n,k}\|_2^2} {\|\mbf{h}_{l_n,k}\|_2^2}, 
\end{equation}
where $\hat{\mbf{h}}_{l_n,k}$ is the reconstructed channel vector and $\mbf{h}_{l_n,k}$ is the original one.

\subsection{Simulation Settings}
 The simulations are carried out for a massive MIMO-OFDM system having $N_t = 64$, antenna spacing of $d=\lambda/15$, carrier wavelength $\lambda$, operating at a carrier frequency of $f_c$ = $2$ GHz. Furthermore, we have a communication bandwidth of $B$ = $20$ MHz, $K = 1$ and $K=3$ UEs, $N = 50$, $P = 500$ test channel vectors, $N_c = 32$ and $M = N_r N_t/2$. For $M = 32$, all the existing and the proposed dictionaries reduce the CSI feedback by 50\% in the uplink. For all the experiments, the subcarrier K-SVD based dictionaries are learned for $N=50$ from each subcarrier.

 For experiments in the multi-UE system, the channels are generated using a Quasi Deterministic Radio Channel Generator (QuaDRiGa)~\cite{Fu_2018_ISAPE_Quadriga,Cerdeira_2014_EUCAP_Quadriga} for the three UEs having velocities of $V = 10$, $15$ and $20$ kmph. For experiments in the single-UE system, the channels are taken from the UE of $V = 20$ kmph. The channel update rate (CUR) considered to generate channels in QuaDRiGa is $10$ ms. The QuaDRiGa simulation platform is recommended by 3GPP (3rd Generation Partnership Project) for designing and simulating wireless communications systems.

The main motivation for the CDL is not only to reduce the CSI feedback but also to reduce the CSI reconstruction NMSE. The benefit of the proposed methods {in terms of} the NMSE performance has to be studied. Hence we have conducted experiments for determining which of the proposed methods will best replace the DFT dictionary and subcarrier K-SVD based dictionaries in both single-UE and multi-UE systems. Since we have considered $N_c=32$, it is not feasible to show the CD performance across all the subcarriers. In the single-UE system, we have considered subcarriers $l=1$ or $8$, since the channel gains of these two subcarriers are relatively low over the period of time. In the multi-UE system, we have considered the first subcarrier $l=1$ of all three UEs to evaluate the NMSE performance as a function of sparsity and compared the proposed methods' CD performance to the DFT and K-SVD dictionaries in the literature.

In OFDM systems, the need for subcarrier K-SVD based dictionaries increases with the number of subcarriers. The FDCHTFs of each subcarrier are considered to be independent in a wideband OFDM system, but this is only realistic for extremely long CIRs. Hence in the proposed system we assume having realistic correlation among subcarriers in the FD. This correlation among the subcarriers is captured by the CD using one of the two proposed methods. Our CDL procedure may also be extended to larger $N_c$.

In all the simulation results, for each subcarrier it can be observed that the NMSE decreases as the sparsity increases. This is because in the sparse vector transformation~\eqref{Eq:spar_1} the sparse vector $\tilde{\mbf{h}}$ picks many columns in the dictionary for higher sparsities, which in turn helps to solve the optimization problem~\eqref{Eq:spar_2} by minimizing the distance between $\mbf{h}$ and $\tilde{\mbf{h}}$. Hence, higher sparsity will improve the reconstruction performance.


\begin{Experiment}  
	
	In this single-UE experiment, we study the NMSE performance across a particular subcarrier of the UE using the CD learned from the proposed methods and existing methods as a function of sparsity. The proposed methods' CD is learned across all the subcarriers of the UE. For this experiment we consider a massive SU-MIMO-OFDM system having $N_c = 32$ subcarriers and a UE having $N_r = 1$ RA and moving with a velocity of 20 kmph.
	
\end{Experiment}


\begin{figure}[hbt!]
	\centering
	\includegraphics[width=3.6in]{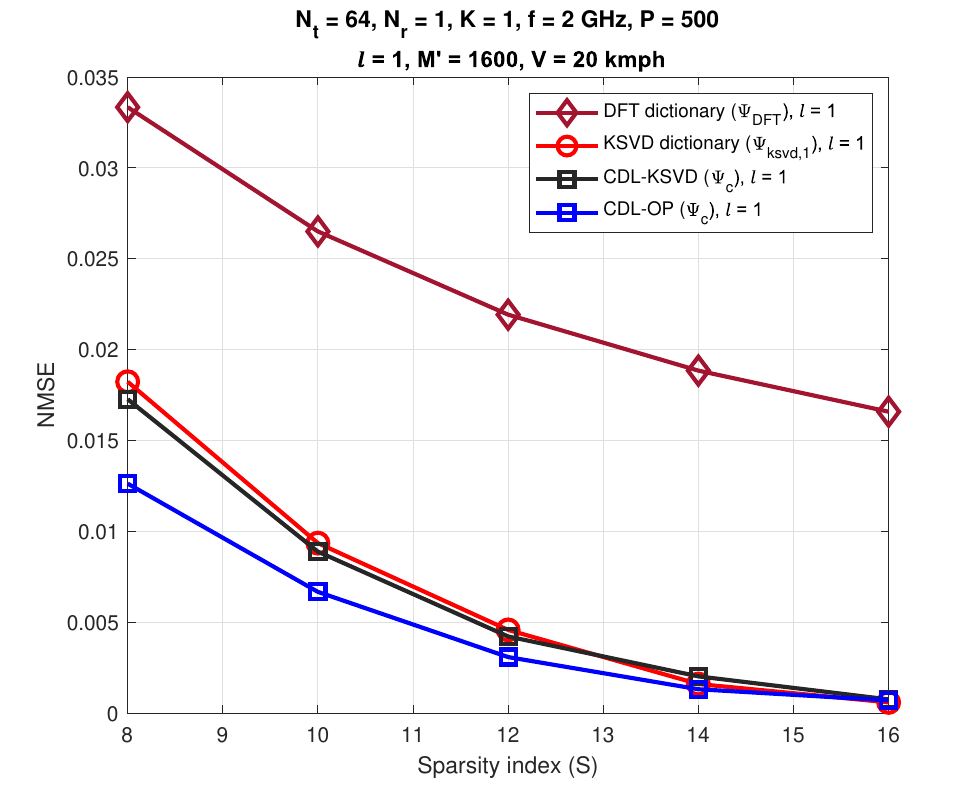}
	\caption{\footnotesize A single-UE system is considered, where the NMSE performance comparison among all the dictionaries is carried out for $N_t$ = 64, $N_r$ = 1 and $N_c$ = 32. Subcarrier 1 is considered for comparison and the UE velocity is 20 kmph.}
	\label{Fig:wb_quad_Su_32subcarr_comparison}
\end{figure}

In Fig. \ref{Fig:wb_quad_Su_32subcarr_comparison}, we consider the subcarrier $l=1$ to study the NMSE performance of all the dictionaries. For a particular sparsity index of $S=8$, the NMSE of the CDL-KSVD dictionary is $1.7 \times 10^{-2}$, of the CDL-OP dictionary is $1.3 \times 10^{-2}$, of the subcarrier K-SVD based dictionary is $1.8 \times 10^{-2}$, and of the DFT dictionary is $3.3 \times 10^{-2}$. The CDL-OP dictionary has the lowest NMSE at $S=8$. For all other sparsities, the CDL-OP and the CDL-KSVD method dictionaries perform similarly, and the NMSE values are close to those of the subcarrier K-SVD dictionary.  All the proposed methods' CD exhibit better performance than the DFT dictionary. 



\begin{figure}[hbt!]
	\centering
	\includegraphics[width=3.6in]{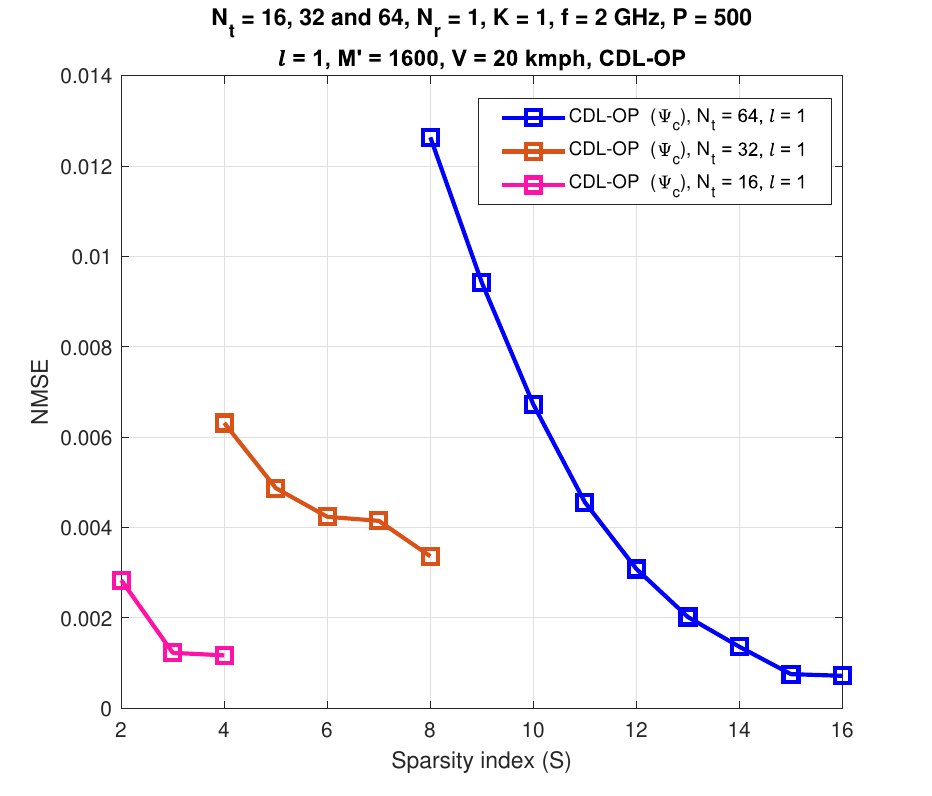}
	\caption{\footnotesize A single-UE system is considered, where the NMSE performance comparison among the CDL-OP dictionaries is carried out for $N_t$ = 16, 32 and 64, $N_r$ = 1 and $N_c$ = 32. Subcarrier 1 is considered for comparison and the UE velocity is 20 kmph.}
	\label{Fig:wb_quad_Su_32_OP_16_32_64}
\end{figure}

In Fig. \ref{Fig:wb_quad_Su_32_OP_16_32_64}, we consider a scenario where the BS has a different number of antennas, namely $N_t$ = 16, 32, and 64. We consider the subcarrier $l=1$ to study the NMSE performance of the CDL-OP dictionaries learned for different $N_t$ values. We observe that the NMSE value increases with the number of antennas at the BS.

    
\begin{Experiment} 
	
   In this single-UE experiment, we study the NMSE performance across different subcarriers of the UE employing the CD learned by the proposed methods and the individual subcarrier K-SVD based dictionaries as a function of sparsity. For this experiment we consider a massive SU-MIMO-OFDM system having $N_c = 32$ subcarriers. { The UE is equipped with $N_r = 1$ (and $2$) RAs and is moving at a velocity of 20 kmph.}

\end{Experiment}


\begin{figure}[hbt!]
	\centering
	\includegraphics[width=3.6in]{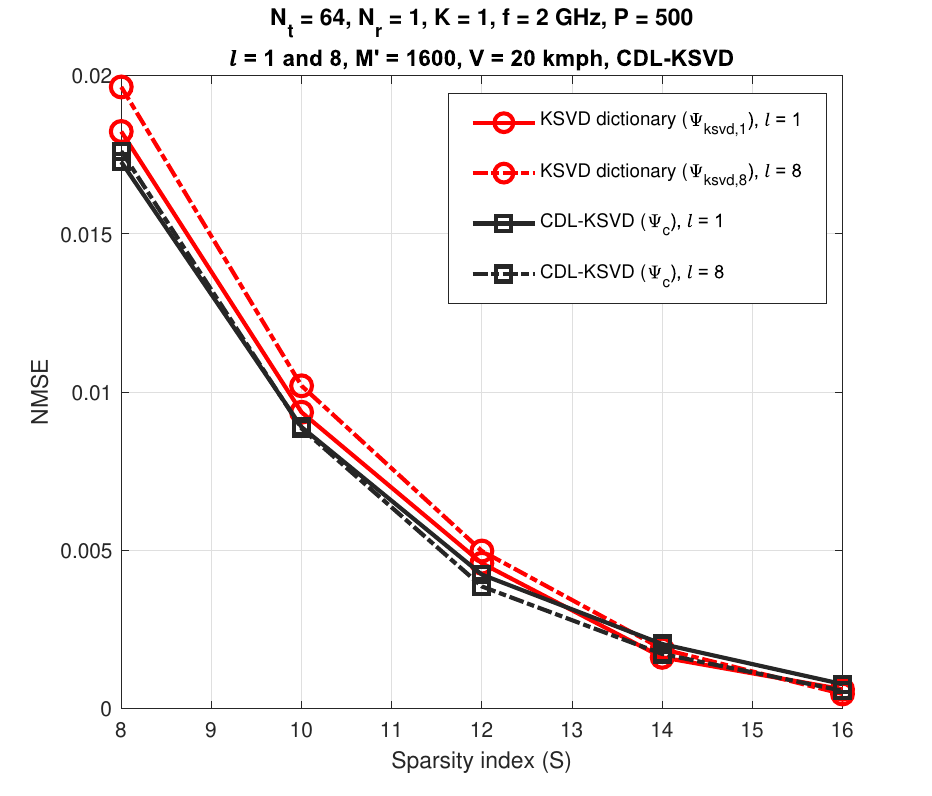}
	\caption{\footnotesize A single-UE system is considered, where the NMSE performance of the K-SVD dictionary and the CDL-KSVD dictionary are used for $N_t$ = 64, $N_r$ = 1 and $N_c$ = 32. Subcarriers 1 and 8 are considered for comparison and the UE velocity is 20 kmph.} 
	\label{Fig:wb_quad_su_32subcarr_ksvd_dics}
\end{figure}


\begin{figure}[hbt!]
	\centering
	\includegraphics[width=3.6in]{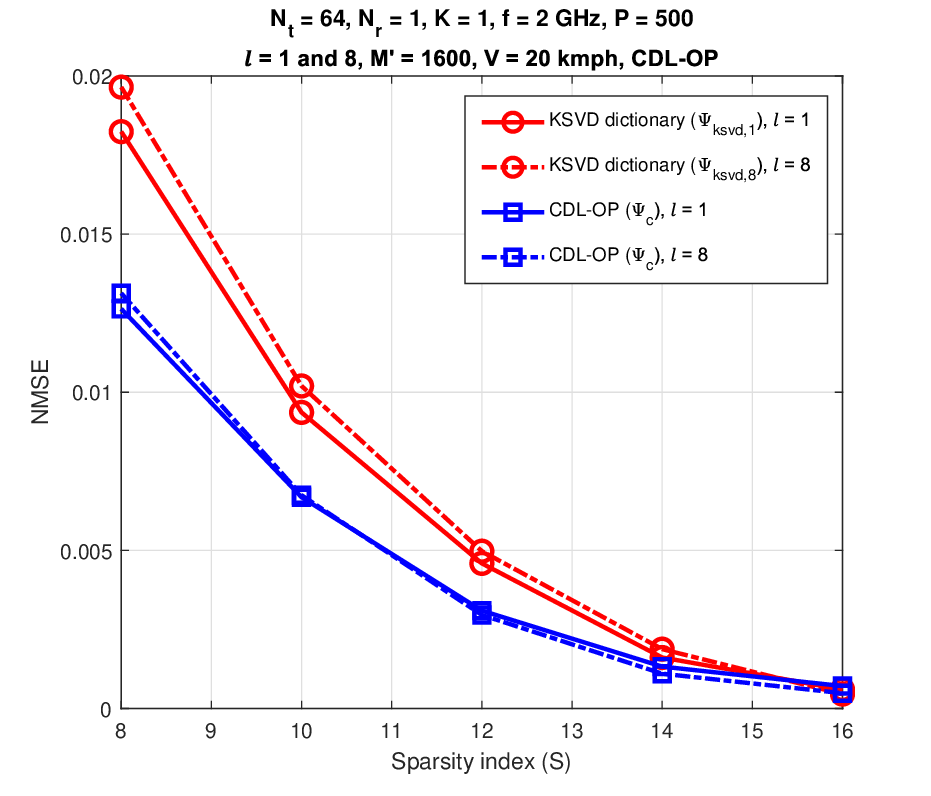}
	\caption{\footnotesize A single-UE system is considered, where the NMSE performance of the K-SVD dictionary and the CDL-OP dictionary are used for for $N_t$ = 64, $N_r$ = 1 and $N_c$ = 32. Subcarriers 1 and 8 are considered for comparison and the UE velocity is 20 kmph.} 
	\label{Fig:wb_quad_su_32subcarr_orth_dics}
\end{figure}

 In Fig. \ref{Fig:wb_quad_su_32subcarr_ksvd_dics}, the $\bols{\Psi}_c$ employed for NMSE characterization is learned from the CDL-KSVD method and in Fig. \ref{Fig:wb_quad_su_32subcarr_orth_dics}, the $\bols{\Psi}_c$ employed for NMSE characterization is learned from the CDL-OP method. Observe from Figs. \ref{Fig:wb_quad_su_32subcarr_ksvd_dics} and \ref{Fig:wb_quad_su_32subcarr_orth_dics}, for subcarriers $l=1$ and $8$ at all the sparsity index values, the NMSE values are close to each other, but the CDL-OP method is the best among all the three methods in terms of the NMSE performance attained.  
 {Both the CDL-KSVD and CDL-OP methods rely on the SVD operation and learn the CD from the channels estimated at the UE. Consequently, the NMSE reconstruction results shown in Figs. \ref{Fig:wb_quad_su_32subcarr_ksvd_dics} and \ref{Fig:wb_quad_su_32subcarr_orth_dics} exhibit a high degree of similarity.}

\begin{figure}[hbt!]
	\centering
	\includegraphics[width=3.6in]{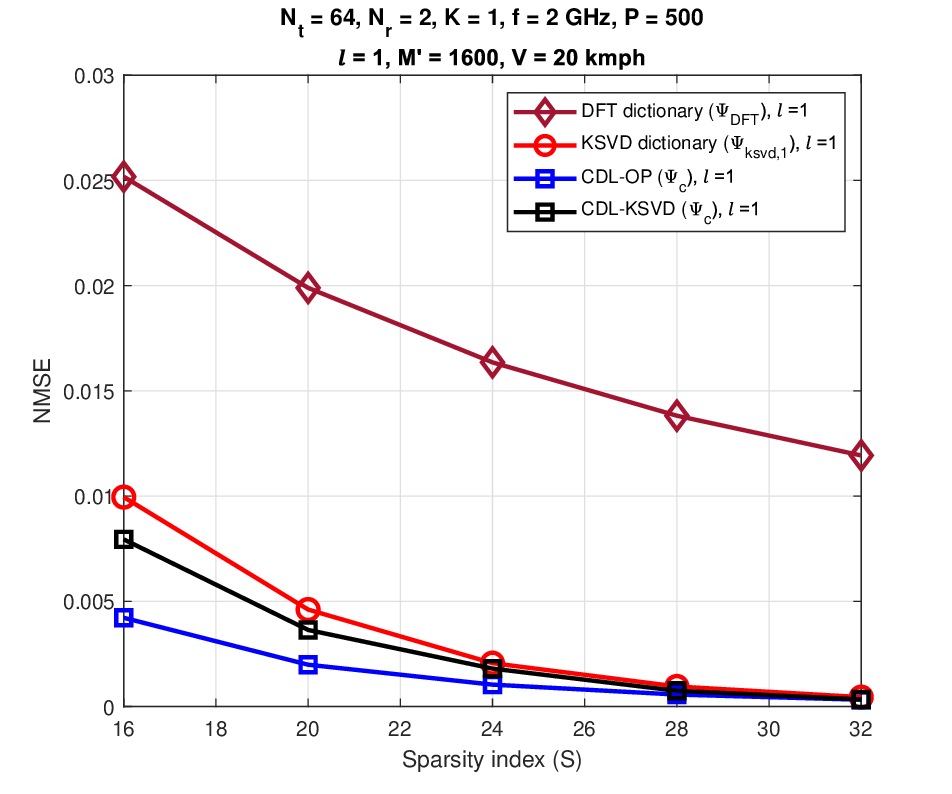}
	\caption{\footnotesize {A single-UE system with multiple UE antennas is considered, where the NMSE performance comparison among all the dictionaries is carried out for $N_t$ = 64, $N_r$ = 2 and $N_c$ = 32. Subcarrier 1 is considered for comparison and the UE velocity is 20 kmph.}}
	\label{Fig:wb_quad_Su_32subcarr_comparison_2rx}
\end{figure}

{We have carried out a simulation and presented the results in Fig. \ref{Fig:wb_quad_Su_32subcarr_comparison_2rx}, where we specifically focus on the scenario where $N_r>1$, indicating the presence of multiple receive antennas. By considering this scenario, we ensure that our analysis is not limited to a specific number of receive antennas. It is observed that all the proposed methods' CD exhibit better NMSE performance than the DFT dictionary.}


\begin{Experiment}  
	
	In this multi-UE experiment, we initially study the NMSE performance of the dictionary generated using the CDL-OP method and existing methods for a particular UE's subcarrier as a function of sparsity. Then we study the NMSE performance of the CDL-OP dictionary across different UEs for a particular subcarrier.
	The CDL-OP dictionary is learned across all the subcarriers and all the $K$ UEs. For the experiment, we consider a massive MU-MIMO-OFDM system having $N_c = 32$ subcarriers and $K = 3$ UEs each associated with $N_r = 1$ RA.

\end{Experiment}


\begin{figure}[hbt!]
	\centering
	\includegraphics[width=3.6in]{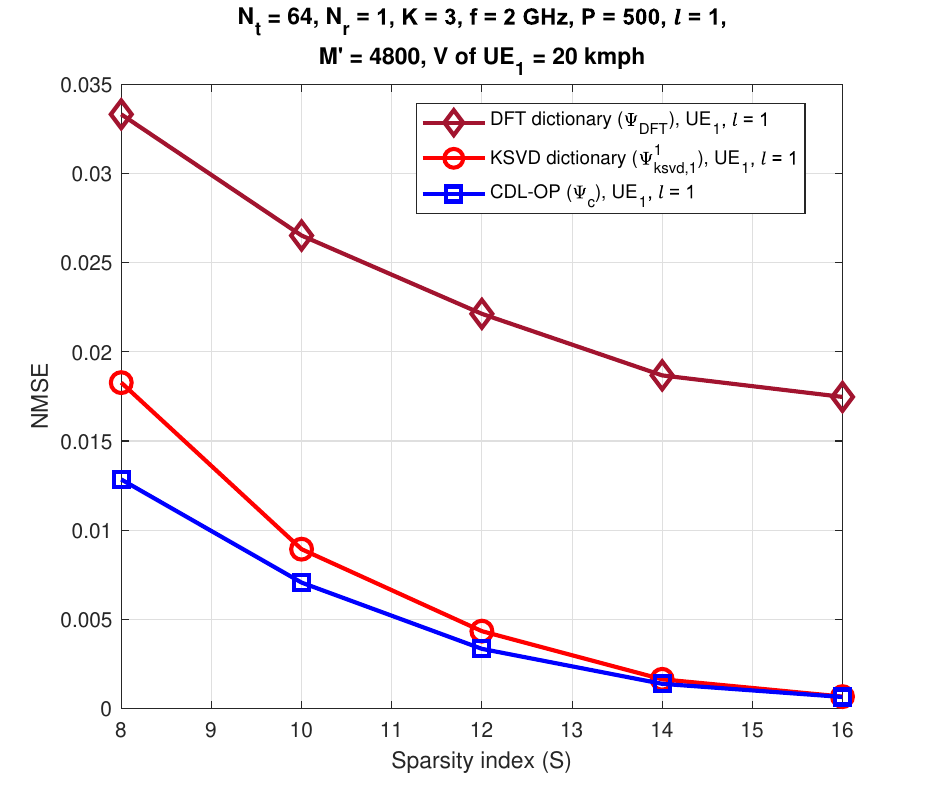}
	\caption{\footnotesize A multi-UE system is considered, where the NMSE performance comparison of the DFT, the K-SVD and the CDL-OP methods  is carried out for  $K$ = 3, $N_t$ = 64, $N_r$ = 1 and $N_c$ = 32. Subcarrier 1 of UE$_1$ is considered for comparison. The velocities of the three UEs are 10, 15, and 20 kmph.}
	\label{Fig:wb_Mu_32subcarr_comparison}
\end{figure}

The wireless channels are generated for three UEs having velocities of $V =$ 10, 15, and 20 kmph using a QuaDRiGa simulator. Let us assume that a UE changes its velocity to that of another UE, but experiences different channel characteristics. In that case, there is no need to learn a new dictionary for that particular UE. The CD has already captured all the three UE channel characteristics, and this procedure can be extended to larger $K$ and $N_c$ values.

In Fig. \ref{Fig:wb_Mu_32subcarr_comparison}, we consider the first UE and subcarrier $l=1$ to study the NMSE performance of the dictionaries from the full set of $N_c=32$ subcarriers and $K=3$ UEs. For a particular sparsity of $S=8$, the NMSE value of the CDL-OP dictionary is $1.3 \times 10^{-2}$, of the subcarrier K-SVD based dictionary is $1.8 \times 10^{-2}$, and of the DFT dictionary is $3.3 \times 10^{-2}$. We observe that all the proposed methods' CD have better NMSE performance than the DFT dictionary. {Among the proposed methods, the CDL-KSVD method exhibits the poorest NMSE performance. Therefore, we are not pursuing this method any further in the simulations.}



\begin{figure}[hbt!]
	\centering
	\includegraphics[width=3.6in]{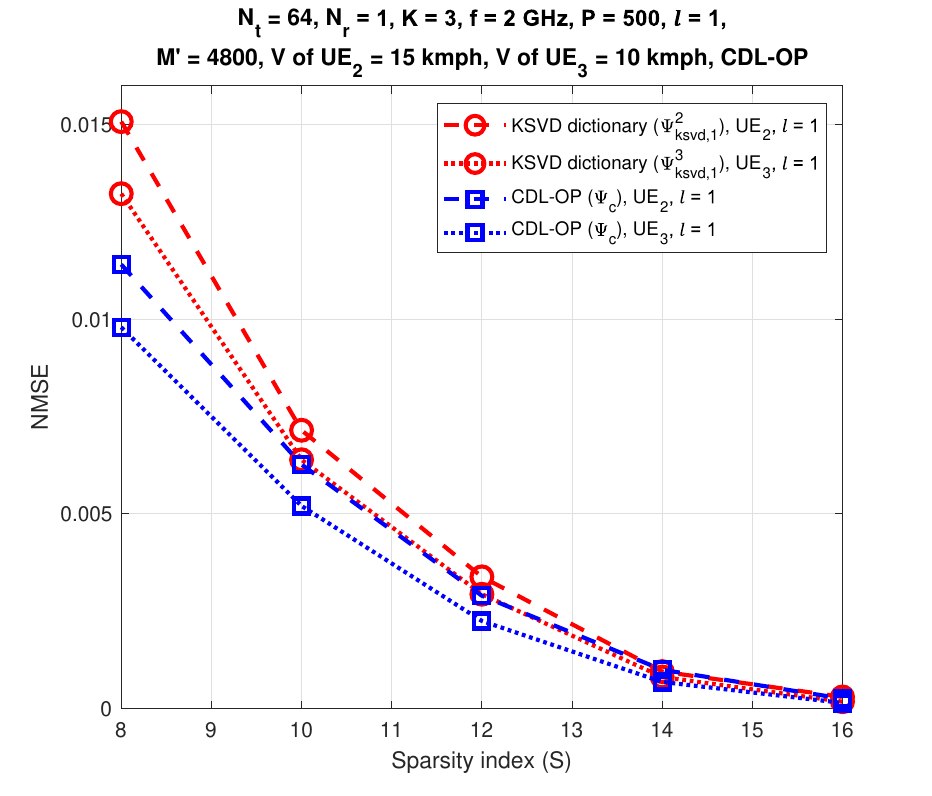}
	\caption{\footnotesize In a multi-UE system, the NMSE performance of the K-SVD dictionary and CDL-OP method's dictionary for $K$ = 3, $N_t$ = 64, $N_r$ = 1 and $N_c$ = 32. Subcarrier 1 of UE$_2$ and UE$_3$ are considered for comparison. Velocities of three UEs are 10, 15, and 20 kmph.}
	\label{Fig:wb_quad_Mu_32subcarr_orth_dics}
\end{figure}

In Fig. \ref{Fig:wb_quad_Mu_32subcarr_orth_dics}, the $\bols{\Psi}_c$ employed for NMSE characterization is learned from the CDL-OP method. Observe from the Fig. \ref{Fig:wb_quad_Mu_32subcarr_orth_dics}, for subcarrier $l=1$ of UE$_2$ and UE$_3$ at almost all the sparsity index values, the NMSE values for the proposed CDL-OP method's dictionary is better than subcarrier K-SVD based dictionaries, because the CDL-OP dictionary is learned from the estimated channels of K-SVD based dictionaries.


\begin{Experiment}  
	In this single-UE experiment, we study the BER performance at the {UE} as a function of the SNR. The data symbols are transmitted from the {BS} to the {UE} in two ways: a) Using the {true channel estimates without compression} and b) Using the {channel estimates with compression} that are obtained from the CDL-OP dictionary.
	
\end{Experiment}


\begin{figure}[hbt!]
	\centering
	\includegraphics[width=3.7in]{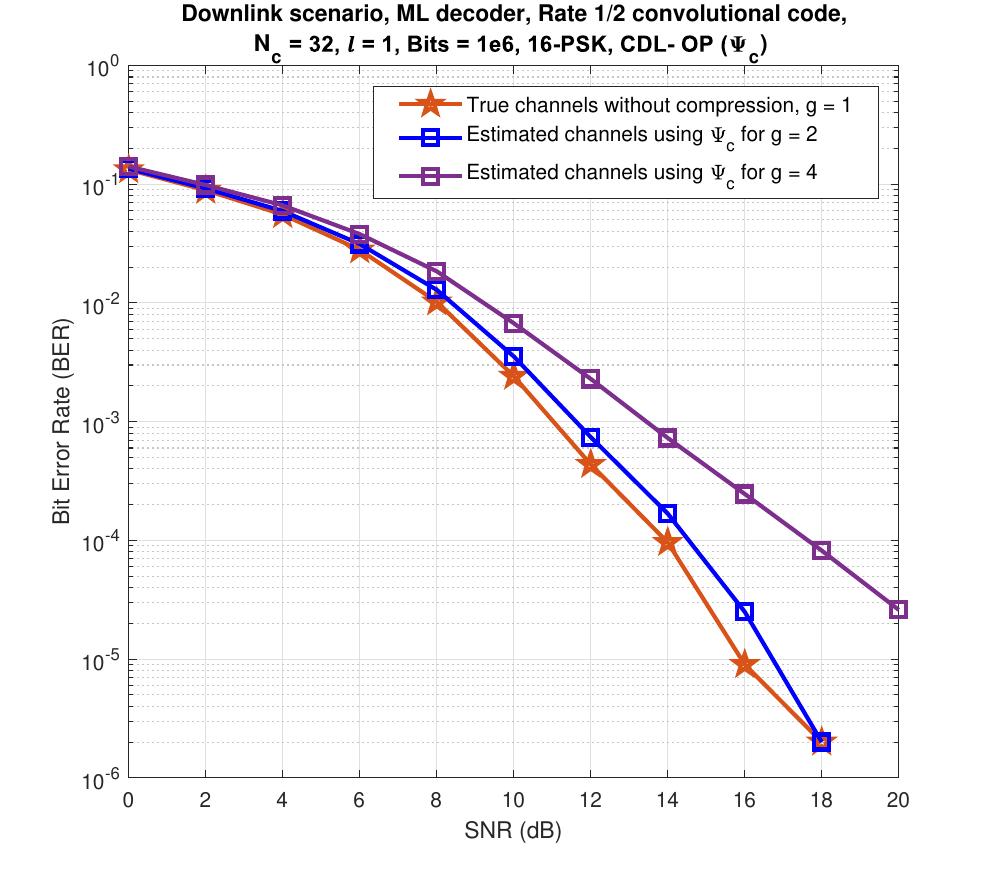}
	\caption{\footnotesize {In a single-UE system with 64 TAs at the BS and a UE with 1 RA, we examine the BER performance of channel estimates using the CDL-OP methods' dictionary. We consider two scenarios: $N_g = N_rN_t/2$ and $N_g = N_rN_t/4$. For comparison, we assume uncompressed channel estimates for subcarrier 1. The UE has a velocity of 20 kmph.}}
\label{Fig:BER_coded_quad_64A_16Psk}
\end{figure}

{To elaborate further on the performance of our proposed
framework, we analyze the BER in a downlink scenario. In Fig. \ref{Fig:BER_coded_quad_64A_16Psk} we evaluate the BER using two sets of channels: the true uncompressed channels and the channels estimated using the CDL-OP dictionary on subcarrier $1$.} {The precoder matrices employed at the BS are denoted as $\mbf{W}^{g}_{} = diag(\frac{{\mbf{h}}_{1}^{H}}{|{\mbf{h}}_{1}|})$ and $\mbf{W}^{g}_{op} = diag(\frac{\hat{\mbf{h}}_{1,op}^{H}}{|\hat{\mbf{h}}_{1,op}|})$, which correspond to the weights obtained from the true channels and to the channels estimated using CDL-OP dictionary, respectively. Let $\mbf{x}$ represents the modulated symbol vector.
Explicitly, in our BER analysis, we harness a half-rate convolutional encoder having the generator sequences of G = [101, 111] and the resultant bits are
16-PSK modulated for generating the symbol vector $\mbf{x}$. The received signal in the case of true uncompressed channels at the UE can be represented as ${y} = \mbf{h}_1^T \mbf{W}^g \mbf{x} + n $, where  $n$ is the additive noise. Similarly the receieved signal for channels estimated
using the CDL-OP dictionary can be represented as ${y_{op}} = \mbf{h}_1^T \mbf{W}^g_{op} \mbf{x} + n $.}

{The CDL-OP dictionaries used in this experiment are learned at a sparsity of 16 for $g=2$ and at a sparsity of 8 for $g=4$. As we increase the compression factor $g$ to reduce the CSI feedback from the UE, we observe a BER performance erosion. The lower bound shown in Fig.~\ref{Fig:BER_coded_quad_64A_16Psk} represents the BER performance when the BS employs uncompressed CSI.
Notably, as depicted in Fig. \ref{Fig:BER_coded_quad_64A_16Psk}, the BER of the UE, recorded for $g=2$ when utilizing the CDL-OP dictionary approaches the lower bound.}

\section{Summary and Conclusions}

 We proposed a novel CDL framework for reducing the FDCHTF feedback and memory requirement of the UE and BS. The framework is more beneficial for the UE, which is usually resource-constrained, and the savings can be significant. For the simulations, the channels are generated using the QuaDRiGa simulator. In a multi-UE system of three UEs and in a single-UE system, all the proposed method's dictionaries have better NMSE performance than the DFT dictionary.  
 The CDL-OP dictionary performs better than the CDL-KSVD dictionary for all the sparsities. Hence the CDL-OP method can be beneficially employed for CDL to compress {the} CSI and improve the CSI reconstruction performance. In terms of the computational complexity, the CDL-OP method requires only {a single} SVD operation to learn the CD, {while the} CDL-KSVD method requires an SVD operation for learning each column in the CD. So the CDL-OP method has {lower} computational complexity than the CDL-KSVD method. {To minimize the impact imposed on the BER performance, it is important to choose an appropriate compression ratio $g$. In our case, a compression ratio of $g=2$ is considered to have a modest impact on the BER performance. By selecting a suitable compression ratio, the system can strike a balance between reducing the amount of feedback, while maintaining a satisfactory BER performance.}

We conclude by highlighting the differences between the multi-UE and single-UE systems as follows:

\begin{enumerate}
 \item In the multi-UE system, the memory is reduced by a factor of $N_c$ at the UE and $KN_c$ at the BS by having only a single CD instead of multiple subcarrier dictionaries. For $N_t = 64$ and $N_c = 32$, the memory required for storing the CD at each UE is reduced by a factor of 32 and the CSI feedback is reduced by a factor of two. 

 \item In the single-UE system, the memory is reduced by a factor of $N_c$ at both the UE and BS, the dictionary feedback is also reduced by a factor of $N_c$  in the uplink. {This} is achieved by sending only {a single} CD instead of $N_c$ subcarrier dictionaries to the BS.  For $N_t = 64$ and $N_c = 32$, the memory is reduced by a factor of 32
 for storing only a CD, and the dictionary feedback is also reduced by a factor of 32. Finally, the CSI feedback is reduced by a factor of two.

 \item In the multi-UE system, the CD ($\bols{\Psi}_{c}$) generated using the CDL-OP has a better NMSE performance to that of the individual subcarrier K-SVD dictionaries $\bols{\Psi}^k_{ksvd,l}$ $\forall l \in [1,N_c], \forall k \in [1,K]$.

 \item In the single-UE system, the CD ($\bols{\Psi}_{c}$) generated using the CDL-OP method has a better NMSE performance to that of the individual subcarrier K-SVD dictionaries $\bols{\Psi}_{ksvd,l}$ $\forall l \in [1,N_c]$.
 
	 
\end{enumerate}

	
\medskip
\bibliography{References}

\begin{thebibliography}{10}
\providecommand{\url}[1]{#1}
\csname url@samestyle\endcsname
\providecommand{\newblock}{\relax}
\providecommand{\bibinfo}[2]{#2}
\providecommand{\BIBentrySTDinterwordspacing}{\spaceskip=0pt\relax}
\providecommand{\BIBentryALTinterwordstretchfactor}{4}
\providecommand{\BIBentryALTinterwordspacing}{\spaceskip=\fontdimen2\font plus
\BIBentryALTinterwordstretchfactor\fontdimen3\font minus
  \fontdimen4\font\relax}
\providecommand{\BIBforeignlanguage}[2]{{%
\expandafter\ifx\csname l@#1\endcsname\relax
\typeout{** WARNING: IEEEtran.bst: No hyphenation pattern has been}%
\typeout{** loaded for the language `#1'. Using the pattern for}%
\typeout{** the default language instead.}%
\else
\language=\csname l@#1\endcsname
\fi
#2}}
\providecommand{\BIBdecl}{\relax}
\BIBdecl

\bibitem{Sanguinetti_2018_IEEETWC_spatial_mux}
E.~Björnson, J.~Hoydis, and L.~Sanguinetti, ``Massive {MIMO} has unlimited
  capacity,'' \emph{IEEE Trans. Wireless Commun.}, vol.~17, no.~1, pp.
  574--590, 2018.

\bibitem{Lajos_2015_IEEECOMSURV_50Years_MIMO}
S.~Yang and L.~Hanzo, ``Fifty years of {MIMO} detection: The road to
  large-scale {MIMO}s,'' \emph{IEEE Commun. Surv. Tutorials.}, vol.~17, no.~4,
  pp. 1941--1988, 2015.

\bibitem{Liu_2019_IEEETWC_MU_Precoding}
R.~Shafin and L.~Liu, ``Multi-cell multi-user massive {FD-MIMO}: Downlink
  precoding and throughput analysis,'' \emph{IEEE Trans. Wireless Commun.},
  vol.~18, no.~1, pp. 487--502, 2019.

\bibitem{Ge_2020_GLOBECOM_Reciprocity}
Z.~Zhong, L.~Fan, and S.~Ge, ``F{DD} massive {MIMO} uplink and downlink channel
  reciprocity properties: Full or partial reciprocity?'' in \emph{Proc. IEEE
  Global Commun. Conf. (GLOBECOM)}, 2020, pp. 1--5.

\bibitem{Lajos_2015_IEEEWCOM_Large_MIMO}
Z.~Zhang, X.~Wang, K.~Long, A.~V. Vasilakos, and L.~Hanzo, ``Large-scale
  {MIMO}-based wireless backhaul in 5{G} networks,'' \emph{IEEE Wireless
  Commun.}, vol.~22, no.~5, pp. 58--66, 2015.

\bibitem{Gadamsetty_2023_NCC_Dictionary}
P.~K. Gadamsetty and K.~V.~S. Hari, ``A fast dictionary learning algorithm for
  {CSI} feedback in massive {MIMO FDD} systems,'' in \emph{Proc. National Conf.
  on Commun. (NCC)}, 2023, pp. 1--6.

\bibitem{ozbek_2020_IEEEVTC_compressive}
S.~S. Yılmaz and B.~Özbek, ``Compressive sensing based low complexity user
  selection for massive {MIMO} systems,'' in \emph{Proc. VTC Spring IEEE 91st
  Veh. Technol. Conf.}, 2020, pp. 1--6.

\bibitem{Mo_2017_IEEETSP_OMP}
J.~Wen, Z.~Zhou, J.~Wang, X.~Tang, and Q.~Mo, ``A sharp condition for exact
  support recovery with orthogonal matching pursuit,'' \emph{IEEE Trans. Signal
  Process.}, vol.~65, no.~6, pp. 1370--1382, 2017.

\bibitem{Fu_2017_IEEETIT_Basis_Pursuit}
X.-J. Liu, S.-T. Xia, and F.-W. Fu, ``Reconstruction guarantee analysis of
  basis pursuit for binary measurement matrices in compressed sensing,''
  \emph{IEEE Trans. Inf. Theory}, vol.~63, no.~5, pp. 2922--2932, 2017.

\bibitem{Xian_2019_IEEESPL_CAMP}
H.~Ge, L.~Wang, J.~Wen, and J.~Xian, ``An {RIP} condition for exact support
  recovery with covariance-assisted matching pursuit,'' \emph{IEEE Signal
  Process. Lett.}, vol.~26, no.~3, pp. 520--524, 2019.

\bibitem{Han_2015_GLOBECOM_basis}
L.~Lu, G.~Y. Li, D.~Qiao, and W.~Han, ``Sparsity-enhancing basis for
  compressive sensing based channel feedback in massive {MIMO} systems,'' in
  \emph{Proc. IEEE Global Commun. Conf. (GLOBECOM)}, 2015, pp. 1--6.

\bibitem{Lajos_2005_BOOK_ofdm_MU}
L.~Hanzo, B.~Choi, T.~Keller \emph{et~al.}, \emph{OFDM and MC-CDMA for
  broadband multi-user communications, WLANs and broadcasting}.\hskip 1em plus
  0.5em minus 0.4em\relax John Wiley \& Sons, 2005.

\bibitem{Chen_2014_VTC_Csi_ofdm}
P.~Cheng and Z.~Chen, ``Multidimensional compressive sensing based analog {CSI}
  feedback for massive {MIMO-OFDM} systems,'' in \emph{Proc. VTC Fall - IEEE
  80th Veh. Technol. Conf.}, 2014, pp. 1--6.

\bibitem{Wang_2021_IEEECOML_RLS_dictionary}
W.~Zeng, Y.~He, B.~Li, and S.~Wang, ``Sparsity learning-based {CSI} feedback
  for {FDD} massive {MIMO} systems,'' \emph{IEEE Wireless Commun. Lett.},
  vol.~10, no.~3, pp. 585--588, 2021.

\bibitem{Li_2016_IEEECOML_Ref_KSVD}
Z.~{Lv} and Y.~{Li}, ``A channel state information feedback algorithm for
  massive {MIMO} systems,'' \emph{IEEE Commun. Lett.}, vol.~20, no.~7, pp.
  1461--1464, 2016.

\bibitem{Jiang_2019_IEEETCOM_Low_rank}
S.~Qiu, D.~Gesbert, D.~Chen, and T.~Jiang, ``A covariance-based hybrid channel
  feedback in {FDD} massive {MIMO} systems,'' \emph{IEEE Trans. Commun.},
  vol.~67, no.~12, pp. 8365--8377, 2019.

\bibitem{Prelcic_2020_IEEETWC_mmwave}
H.~Xie and N.~González-Prelcic, ``Dictionary learning for channel estimation
  in hybrid frequency-selective mm{W}ave {MIMO} systems,'' \emph{IEEE Trans.
  Wireless Commun.}, vol.~19, no.~11, pp. 7407--7422, 2020.

\bibitem{Lau_2022_GLOBECOM_DL}
Y.~Zhao, Y.~Teng, A.~Liu, and V.~Lau, ``Two-timescale joint {UL/DL} dictionary
  learning and channel estimation in massive {MIMO} systems,'' in \emph{Proc.
  IEEE Global Commun. Conf. (GLOBECOM)}, 2022, pp. 5408--5413.

\bibitem{Savazzi_2020_IEEEACC_CSI_Feedback}
F.~Kulsoom, A.~Vizziello, H.~N. Chaudhry, and P.~Savazzi, ``Joint sparse
  channel recovery with quantized feedback for multi-user massive {MIMO}
  systems,'' \emph{IEEE Access}, vol.~8, pp. 11\,046--11\,060, 2020.

\bibitem{Shim_2015_IEEETCOM_Antenna_group}
B.~Lee, J.~Choi, J.-Y. Seol, D.~J. Love, and B.~Shim, ``Antenna grouping based
  feedback compression for {FDD}-based massive {MIMO} systems,'' \emph{IEEE
  Trans. Commun.}, vol.~63, no.~9, pp. 3261--3274, 2015.

\bibitem{Heath_2018_IEEETCOM_AOD_codebook}
W.~Shen, L.~Dai, B.~Shim, Z.~Wang, and R.~W. Heath, ``Channel feedback based on
  {A}o{D}-adaptive subspace codebook in {FDD} massive {MIMO} systems,''
  \emph{IEEE Trans. Commun.}, vol.~66, no.~11, pp. 5235--5248, 2018.

\bibitem{Zhao_2021_IEEETVT_AOD_codebook}
G.~Huang, A.~Liu, and M.-J. Zhao, ``Two-stage adaptive and compressed {CSI}
  feedback for {FDD} massive {MIMO},'' \emph{IEEE Trans. Veh. Technol.},
  vol.~70, no.~9, pp. 9602--9606, 2021.

\bibitem{Zhao_2021_IEEECOML_Low_rank}
Z.~Wei, H.~Li, H.~Liu, B.~Li, and C.~Zhao, ``Randomized low-rank approximation
  based massive {MIMO} {CSI} compression,'' \emph{IEEE Commun. Lett.}, vol.~25,
  no.~6, pp. 2004--2008, 2021.

\bibitem{LI_2019_IEEEACC_DL_CSI}
Y.~Liao, H.~Yao, Y.~Hua, and C.~Li, ``C{SI} feedback based on deep learning for
  massive {MIMO} systems,'' \emph{IEEE Access}, vol.~7, pp. 86\,810--86\,820,
  2019.

\bibitem{Li_2019_IEEEWCL_CsiNet_LSTM}
T.~Wang, C.-K. Wen, S.~Jin, and G.~Y. Li, ``Deep learning-based {CSI} feedback
  approach for time-varying massive {MIMO} channels,'' \emph{IEEE Wireless
  Commun. Lett.}, vol.~8, no.~2, pp. 416--419, 2019.

\bibitem{Li_2020_IEEECOML_DL_DNNet}
H.~Ye, F.~Gao, J.~Qian, H.~Wang, and G.~Y. Li, ``Deep learning-based denoise
  network for {CSI} feedback in {FDD} massive {MIMO} systems,'' \emph{IEEE
  Commun. Lett.}, vol.~24, no.~8, pp. 1742--1746, 2020.

\bibitem{Li_2020_IEEETVT_DL_ReNet}
P.~Liang, J.~Fan, W.~Shen, Z.~Qin, and G.~Y. Li, ``Deep learning and
  compressive sensing-based {CSI} feedback in {FDD} massive {MIMO} systems,''
  \emph{IEEE Trans. Veh. Technol.}, vol.~69, no.~8, pp. 9217--9222, 2020.

\bibitem{Nallanathan_2022_IEEETVT_DCRNet}
S.~Tang, J.~Xia, L.~Fan, X.~Lei, W.~Xu, and A.~Nallanathan, ``Dilated
  convolution based {CSI} feedback compression for massive {MIMO} systems,''
  \emph{IEEE Trans. Veh. Technol.}, vol.~71, no.~10, pp. 11\,216--11\,221,
  2022.

\bibitem{Deyu_2023_IEEETWC_Deepunfold}
Z.~Hu, G.~Liu, Q.~Xie, J.~Xue, D.~Meng, and D.~Gündüz, ``A learnable
  optimization and regularization approach to massive {MIMO CSI} feedback,''
  \emph{IEEE Trans. Wireless Commun.}, pp. 1--1, 2023.

\bibitem{Jin_2023_IEEEWCL_Deepunfold_bit}
Z.~Cao, J.~Guo, C.-K. Wen, and S.~Jin, ``Deep-unfolding-based bit-level {CSI}
  feedback in massive {MIMO} systems,'' \emph{IEEE Wireless Commun. Lett.},
  vol.~12, no.~2, pp. 371--375, 2023.

\bibitem{Ting_2012_WCNC_compressive}
P.-H. Kuo, H.~T. Kung, and P.-A. Ting, ``Compressive sensing based channel
  feedback protocols for spatially-correlated massive antenna arrays,'' in
  \emph{Proc. IEEE Wireless Commun. Netw. Conf. (WCNC)}, 2012, pp. 492--497.

\bibitem{Vetterli_2012_IEEETCOM_Joint_sparsity}
Y.~Barbotin, A.~Hormati, S.~Rangan, and M.~Vetterli, ``Estimation of sparse
  {MIMO} channels with common support,'' \emph{IEEE Trans. Commun.}, vol.~60,
  no.~12, pp. 3705--3716, 2012.

\bibitem{Li_2016_ICASSP_SBL}
L.~Yang, J.~Fang, and H.~Li, ``Sparse {B}ayesian dictionary learning with a
  {G}aussian hierarchical model,'' in \emph{Proc. IEEE Int. Conf. Acoust.,
  Speech Signal Process. (ICASSP)}, 2016, pp. 2564--2568.

\bibitem{Cui_2018_IEEEWCL_Data_CL}
K.~Zhang, E.~L. Xu, H.~Zhang, Z.~Feng, and S.~Cui, ``Data driven automatic
  modulation classification via dictionary learning,'' \emph{IEEE Wireless
  Commun. Lett.}, vol.~7, no.~4, pp. 586--589, 2018.

\bibitem{Arslan_2019_IWCMC_DL_Beamspace}
M.~Nazzal, M.~A. Aygül, A.~Görçin, and H.~Arslan, ``Dictionary
  learning-based beamspace channel estimation in millimeter-wave massive {MIMO}
  systems with a lens antenna array,'' in \emph{Proc. 15th Int. Wireless
  Commun. \& Mobile Computing Conf. (IWCMC)}, 2019, pp. 20--25.

\bibitem{Bruckstein_2006_IEEETSP_Ksvdalgo}
M.~{Aharon}, M.~{Elad}, and A.~{Bruckstein}, ``K-{SVD}: An algorithm for
  designing overcomplete dictionaries for sparse representation,'' \emph{IEEE
  Trans. Signal Process.}, vol.~54, no.~11, pp. 4311--4322, 2006.

\bibitem{Dijksterhuis_2004_OxUni_procrustes}
J.~C. Gower, G.~B. Dijksterhuis \emph{et~al.}, \emph{Procrustes
  problems}.\hskip 1em plus 0.5em minus 0.4em\relax Oxford University Press on
  Demand, 2004, vol.~30.

\bibitem{Jin_2018_IEEEWCL_DL_CsiNet}
C.-K. Wen, W.-T. Shih, and S.~Jin, ``Deep learning for massive {MIMO} {CSI}
  feedback,'' \emph{IEEE Wireless Commun. Lett.}, vol.~7, no.~5, pp. 748--751,
  2018.

\bibitem{Albani_2015_IEEETAP_fading_chan_Bessel}
M.~Ettorre, S.~C. Pavone, M.~Casaletti, and M.~Albani, ``Experimental
  validation of {B}essel beam generation using an inward hankel aperture
  distribution,'' \emph{IEEE Trans. Antennas and Propagation}, vol.~63, no.~6,
  pp. 2539--2544, 2015.

\bibitem{Fu_2018_ISAPE_Quadriga}
Z.~F. et~al., ``Channel simulation and validation by {Q}uadriga for suburban
  microcells under 6 {GH}z,'' in \emph{Proc. 12th Int. Symp. on Antennas,
  Propagation and EM Theory (ISAPE)}, 2018, pp. 1--4.

\bibitem{Cerdeira_2014_EUCAP_Quadriga}
F.~Burkhardt, S.~Jaeckel, E.~Eberlein, and R.~Prieto-Cerdeira, ``Quadriga: A
  {MIMO} channel model for land mobile satellite,'' in \emph{Proc. The 8th
  European Conf. on Antennas and Propagation (EuCAP 2014)}, 2014, pp.
  1274--1278.

\end{thebibliography}
\bibliographystyle{IEEEtran}

\end{document}